\documentclass{SciPost}
\usepackage{physics}
\usepackage{algorithm}
\usepackage{algorithmic}
\usepackage{minted}
\usepackage{xcolor}
\usepackage{textcomp}
\usepackage{tikz, pgfplots,tikz-cd}
\usetikzlibrary{calc,decorations.markings,decorations.pathmorphing,shapes}
\usepackage{cqmacros}
\usepackage{subcaption}
\usetikzlibrary{matrix}
\usepackage{amsthm}

\hypersetup{
    colorlinks,
    linkcolor={red!50!black},
    citecolor={blue!50!black},
    urlcolor={blue!80!black}
}

\usepackage[bitstream-charter]{mathdesign}
\DeclareSymbolFont{usualmathcal}{OMS}{cmsy}{m}{n}
\DeclareSymbolFontAlphabet{\mathcal}{usualmathcal}

\fancypagestyle{SPstyle}{
\fancyhf{}
\lhead{\colorbox{scipostblue}{\bf \color{white} ~SciPost Physics Codebases }}
\rhead{{\bf \color{scipostdeepblue} ~Submission }}

\fancyfoot[C]{\textbf{\thepage}}
}

\begin{document}

\pagestyle{SPstyle}
\begin{center}{\Large \textbf{\color{scipostdeepblue}{
A Practical Introduction to Tensor Network Renormalization\\
with TNRKit.jl
}}}\end{center}

\begin{center}\textbf{
Victor Vanthilt\textsuperscript{$1,\star$},
Adwait Naravane\textsuperscript{1},
Chenqi Meng\textsuperscript{2}
Atsushi Ueda\textsuperscript{1}
}\end{center}

\begin{center}
Department of Physics and Astronomy, Ghent University, Krijgslaan 299, 9000 Gent, Belgium\textsuperscript{1}\\
Department of Physics, The Chinese University of Hong Kong, Sha Tin, New Territories, Hong Kong, China\textsuperscript{2}\\
$\star$ \href{mailto:victor.vanthilt@ugent.be}{\small victor.vanthilt@ugent.be}\,\quad
\end{center}

\section*{\color{scipostdeepblue}{Abstract}}
\textbf{\boldmath{We present TNRKit.jl, an open-source Julia package for Tensor Network Renormalization (TNR) of two- and three-dimensional classical statistical models and Euclidean lattice field theories. Built on top of TensorKit.jl\cite{tensorkit}, it provides a symmetry-aware framework for constructing tensor-network representations of partition functions and coarse-graining them using methods such as TRG, HOTRG, and LoopTNR. Beyond thermodynamic quantities, the package enables the extraction of universal conformal data -- including scaling dimensions and the central charge -- directly from fixed-point tensors. TNRKit.jl is designed with both usability and extensibility in mind, offering a practical platform for applying, benchmarking, and developing modern tensor renormalization algorithms. This paper also serves as a self-contained introduction to the TNR framework.}}

\vspace{\baselineskip}

\noindent\textcolor{white!90!black}{%
\fbox{\parbox{0.975\linewidth}{%
\textcolor{white!40!black}{\begin{tabular}{lr}%
  \begin{minipage}{0.6\textwidth}%
    {\small Copyright attribution to authors. \newline
    This work is a submission to SciPost Physics Codebases. \newline
    License information to appear upon publication. \newline
    Publication information to appear upon publication.}
  \end{minipage} & \begin{minipage}{0.4\textwidth}
    {\small Received Date \newline Accepted Date \newline Published Date}%
  \end{minipage}
\end{tabular}}
}}
}

\vspace{10pt}
\noindent\rule{\textwidth}{1pt}
\tableofcontents
\noindent\rule{\textwidth}{1pt}
\vspace{10pt}

\section{Introduction}
\label{sec:intro}
\subsection{Preface}
The last three decades have been golden years for the tensor network community. In one dimension in particular, the density matrix renormalization group (DMRG) algorithm~\cite{White_1992,dmrg2} has changed the landscape of quantum simulation beyond recognition. Thirty years ago, it was state of the art to compute the Haldane gap on a Windows 95 machine with 16 MB of RAM. These were in the days when turning on a computer was an act of faith: you could make breakfast and still return before it had finished waking up. At the time, this was thus a remarkable achievement. After all, when Haldane proposed in 1983 that integer-spin antiferromagnetic chains should be gapped, the claim was bold enough to become known as the Haldane conjecture~\cite{Haldane_1983}. By the time DMRG came along, the existence of the gap was already strongly supported, but DMRG turned its computation from a delicate numerical feat into something routine. Today, somewhat sadly for those of us with a taste for nostalgia, the same calculation takes less than a second on a laptop and reaches six-digit precision without breaking a sweat. There is barely enough time left to make breakfast.

These advances owe a great deal to open-source software such as TensorKit~\cite{tensorkit}, MPSKit~\cite{MPSKit}, ITensor~\cite{itensor}, and Tenpy~\cite{tenpy}. Gone are the days when one had to spend years learning FORTRAN, several more months implementing DMRG by hand, and a few miserable weeks discovering that the fatal bug was sitting in the very first line of the code all along. Thanks to accessible, well-maintained libraries, the barrier to simulating quantum systems is now so low that one no longer needs to reimplement every core algorithm from scratch. And that is exactly as it should be: it leaves us free to ask the questions that are actually deep and interesting.

This paper shares the same spirit. We aim to make tensor network renormalization (TNR) equally accessible, lowering the barrier to its application and further development. To our knowledge, TNRKit.jl is the first comprehensive, publicly available package dedicated to TNR methods.

TNR~\cite{Gu_2009,Looptnr_2017,Hauru_2016,TNR_evenbly_2015,NNR_2024,tnrplus_2017,thermal&globaltnr,Evenbly:2007hxg,GILT} is a branch of tensor network methods that looks at the physics from a slightly different viewpoint. DMRG studies low-energy physics through the Hamiltonian formalism, while TNR approaches the same questions through the path-integral, or partition-function formalism. Each has its own advantages. In particular, TNR often offers a more natural setting for lattice gauge theories and other quantum field theories. At heart, however, they share the same ambition: to capture the low-energy physics of many-body systems.

But why do we care so much about low energy in the first place? In the Hamiltonian formalism, the answer is clear enough. Many systems in condensed matter physics occur in nature in -- or near -- their ground state. Important observables, such as magnetization and correlation functions, are then computed as the expectation values in the ground state, the state with the lowest energy. Similarly, the dominant transport properties are governed by the low-energy part of the dispersion relation. It is therefore the low-energy sector that carries the physics that we usually care most about.  

In the path-integral formalism, however, the importance of low-energy physics is less obvious at first sight. Here, the central idea is the renormalization group (RG)~\cite{Wilson_RG,Wilson_RG1,Wilson_RG2, Kadanoff_review, Kadanoffising}. RG is a way of classifying phases by asking how a system looks when viewed on larger and larger length scales. In the thermodynamic limit, short-distance thermal and quantum fluctuations become irrelevant. The underlying philosophy is actually very simple.

Suppose your system has a correlation length $\xi = 10$ lattice sites. Now, change the parameters slightly so that $\xi = 9.999$, and ask whether you have crossed into a different phase. Probably not. On the scale of a system containing ten million sites, the distinction between 10 and 9.999 is microscopic to the point of irrelevance. Both are effectively zero compared to the size of the system. The renormalization group makes this intuition concrete by \textit{coarse-graining} the system step by step. Instead of leaping directly to the thermodynamic limit, one follows how the theory changes as the system is repeatedly rescaled. If, for instance, the linear size doubles at each step, then the relative correlation length as a fraction of the system size is cut in half each time. Eventually it shrinks to zero, and the theory flows to what is known as a fixed point. The classification of phases is then reduced to the study of theories with zero correlation length, which is a much simpler task than dealing with the original model. Since energy is inversely related to length scale, this is just another way to say that low-energy physics is important.

There is, however, one important exception to the argument above: criticality. At a critical point, the correlation length is not small compared to the system size at all; it is as large as the system itself. The RG flow then lands not on a theory with zero correlation length, but on one with infinite correlation length. Such theories are perfectly valid fixed points, and are called critical fixed points. In two dimensions, many continuous phase transitions enjoy an emergent conformal symmetry, with their universal properties governed by a conformal field theory (CFT)~\cite{Polchinsk_1987dy,BPZ,DiFrancesco:1997nk,cardy_1986}. This is what makes CFT so powerful: it determines the universality class of the phase transition in full. The snag is that conventional numerical methods often struggle to extract the complete conformal data. TNR, by contrast, can do this with striking efficiency.

This review is divided into three parts. In the remainder of Sec.~\ref{sec:intro}, we begin by reviewing the importance of the partition function through the example of the two-dimensional Ising model. We then explain how partition functions can be encoded in the language of tensor networks. To make the most of our package, we also introduce the symmetry-preserving construction of tensor networks. This gives rise to block-diagonal tensors and leads to a significant speedup in practice. In Sec.~\ref{sec:tnr}, we introduce tensor-based RG schemes and some of their more sophisticated descendants. In Sec.~\ref{sec:cft}, we review how universal information can be extracted from critical models. To keep the discussion self-contained, we also review the underlying conformal field theory and explain how it enters in the context of TNR. We conclude with some benchmarks that compare some of the TRG and TNR methods provided by TNRKit in Section \ref{sec:benchmarks}.

The Julia package is available at \url{https://github.com/QuantumKitHub/TNRKit.jl}.

\subsection{Importance of partition functions}
The partition function $Z$ is the core object of interest in statistical physics: it can be interpreted as a generating function of the system's configurations, encoding its full thermodynamic information:
\begin{equation}
    Z = \sum_{\{\sigma\}} e^{-\beta H(\{\sigma\})},
    \label{eq:partition_function}
\end{equation}
where the sum runs over all configurations of the system. The normalised Boltzmann weights $e^{-\beta H(\{\sigma\})}/Z$ can be interpreted as a probability distribution for the configurations. Any observable $\mathcal{O}$ is recovered as a thermal expectation value,
\begin{equation}
    \langle \mathcal{O} \rangle = \frac{1}{Z} \sum_{\{\sigma\}} \mathcal{O}(\{\sigma\})\, e^{-\beta H(\{\sigma\})}
\end{equation}

while thermodynamic quantities such as the Free Energy $F = -\frac{1}{\beta}\ln Z$, the magnetisation, the specific heat and the magnetic susceptibility follow from derivatives of $\ln Z$ with respect to $\beta$ or an external field.

A paradigmatic example is the classical two-dimensional ferromagnetic Ising model on a square lattice. Despite its simplicity, the model exhibits a non-trivial phase structure associated with spontaneous
$\mathbb{Z}_2$ symmetry breaking, admits an exact solution~\cite{Onsager1944}, and
therefore serves as an ideal benchmark for tensor network renormalization methods. 

The partition function for the classical two-dimensional ferromagnetic Ising model on a Square lattice is: 
\begin{equation}
    Z = \sum_{\{\sigma = \pm 1\}} \exp\!\left(\beta \left\{\sum_{(x,\hat{\mu})} \sigma_x \,
    \sigma_{x+\hat{\mu}} + h \sigma_x \right\}\right)
    = \sum_{\{\sigma = \pm 1\}} \prod_{(x,\hat{\mu})} e^{\beta \sigma_x \sigma_{x+\hat{\mu}} + \beta h \sigma_x}.
\end{equation}
Here, $(x, \hat{\mu})$ represents a link from site $x$ to site $x + \hat{\mu}$. Notice that the Hamiltonian \\ $H = -(\sum_{(x, \hat{\mu})} \sigma_x \sigma_{x + \hat{\mu}} + h\sigma_x)$  has a global $\mathbb{Z}_2$ spin-flip symmetry when the external field $h = 0$. It remains invariant under $\sigma_i \rightarrow - \sigma_i \  \forall i$. The magnetisation per site
\begin{equation}
    \langle m\rangle = \frac{1}{V}\frac{\partial \ln Z}{\partial (\beta h)}
\end{equation}
serves as the order parameter of the symmetry; it vanishes in the symmetric high temperature phase and gains a finite non-zero value in the low temperature phase where the discrete $\mathbb{Z}_2$ symmetry is spontaneously broken. The two phases are separated by a second-order phase transition at $\beta_c = \frac{1}{2}\ln(1+\sqrt{2})$~\cite{Onsager1944, Kadanoffising}, where the correlation length diverges and the system exhibits conformal invariance belonging to the universality class of the Ising CFT, characterized by its central charge \(c = \frac{1}{2}\).

\subsection{Tensor network representation of the two-dimensional Ising model}
\begin{figure}[bt]
    \centering
    \includegraphics[page = 2, width=0.8\linewidth]{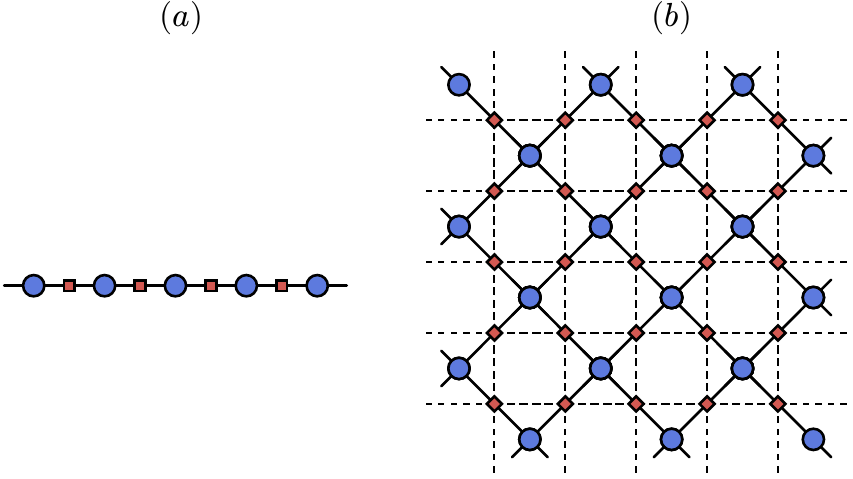}
    \caption{$(a)$ Matrix representation of the 1D partition function. $(b)$ Checkerboard encoding of a 2D partition function. The blue circles and red squares denote the Boltzmann tensors and the spin degrees of freedom to be traced out, respectively.}
    \label{fig:checkerboard}
\end{figure}
At first glance, Eq.~\eqref{eq:partition_function} looks unusable: it contains a sum over an infinitely large number of configurations $\{\sigma\}$. For a generic model, there exists no closed-form solution to calculate it exactly. To make the problem tractable, we first translate it into the language of tensor networks \cite{bridgeman2017hand, cuiper2025houches}, which is the subject of this section, and then later use TRG and TNR methods to approximately -- but accurately -- evaluate this tensor network. If both of these steps are implemented correctly, one can just press run, leave the machine to do the hard work, and head off to a beachside cafe for some crêpes. Let us now explain how to earn this luxurious life.

We begin with the partition function of the classical one-dimensional Ising model with periodic boundary conditions. It can be written as a product of transfer matrices,

\begin{equation}
    Z = \mathrm{Tr}\left(\prod_{i=1}^L M\right),
\end{equation}\\
with
\begin{equation}
    M = \begin{pmatrix}
        e^{\beta} & e^{-\beta}\\
        e^{-\beta} & e^\beta
    \end{pmatrix}\label{eq:M-matrix}.
\end{equation}
The matrix indices correspond to the original spin degrees of freedom. The matrix $M$ is designed such that it assigns the Boltzmann weight $e^{\beta}$ to aligned neighbouring spins and $e^{-\beta}$ to anti-aligned ones. Then, we find that the full partition function under periodic boundary conditions is 
\begin{align*}
    Z &= \sum_{\{\sigma\}} e^{\beta\sum_j \sigma_j\sigma_{j+1}},\\
    &= \Tr \left(M^L\right) = (2\cosh(\beta))^L + (2\sinh(\beta))^L.
\end{align*}
The point to notice is that the sum over spin configurations in the first expression, is replaced by a sum over matrix indices. This is a first example of a tensor-network encoding that we refer to as the \emph{checkerboard construction}. The construction is illustrated in a tensor-network diagram in Fig.~\ref{fig:checkerboard} $(a)$: The blue circles denote the matrices $M$, while the red squares denote the spin indices that are being summed over. 

With this picture in mind, the generalization to two dimensions is fairly straightforward, as shown in panel $(b)$. The partition function of the two-dimensional classical Ising model is encoded as the contraction of four-leg tensors on a square lattice. The tensor itself -- indicated in blue in Figure \ref{fig:checkerboard} (b) -- is quite simple: it is a product of the four Boltzmann weights from the surrounding spins.

Notice that in this \textit{checkerboard construction}, the spins are placed on the edges of the network, and the tensors on the vertices. In the upcoming section \ref{sec:revisitingtheisingmodel}, it will be the other way around.

The checkerboard construction runs into trouble when the model has continuous spin variables, as in the XY model: the dimension of the tensor indices, or the bond dimension, is nothing but the spin degrees of freedom being summed over. For XY spins, this number is infinite, and the construction becomes impractical. Fortunately, there is another route, known as the character expansion, which neatly sidesteps this problem. In this approach, one introduces new variables on the edges of the original lattice. These variables live in the basis of irreducible representations of the symmetry of the model, allowing one to encode the model efficiently by making its symmetry manifest from the start.

\subsection{Character expansion}
\label{sec:Isingmodelinitialtensor}
\begin{figure}[tb]
    \centering
    \includegraphics[page = 19, width=\linewidth]{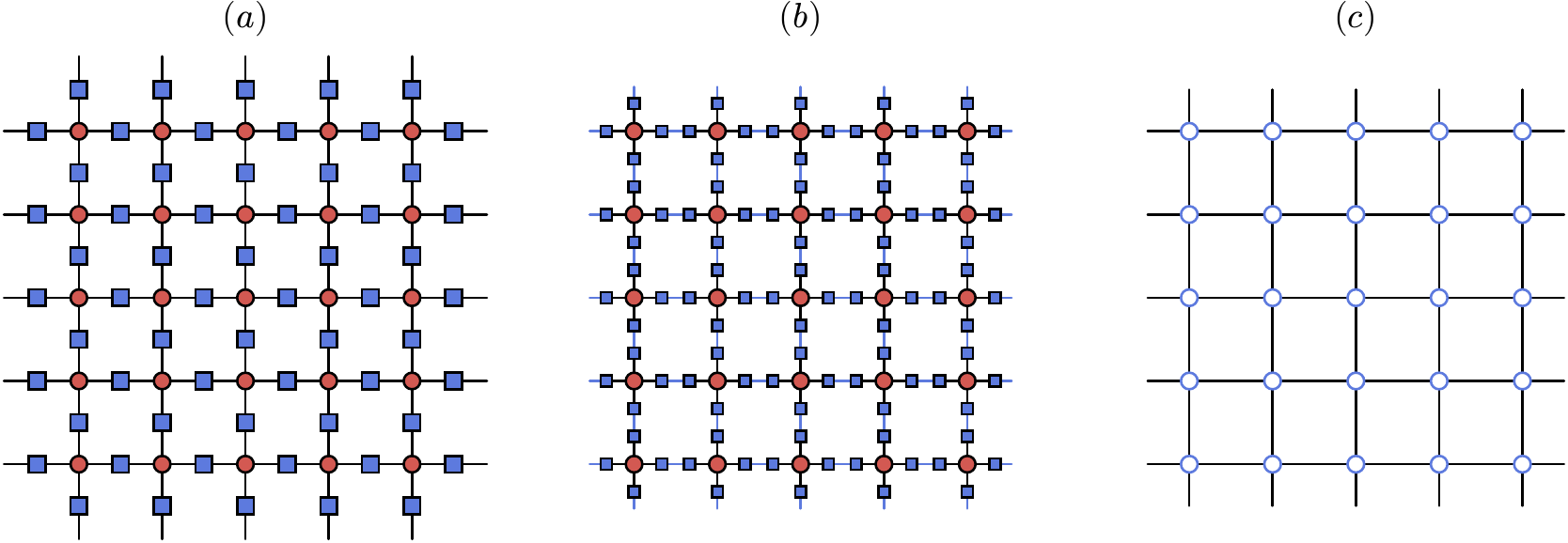}
    \caption{Schematic illustration of constructing the initial tensors via character expansion. $(a)$ The original spins are placed on the vertices, while the Boltzmann matrices live on the edges. $(b)$ Each Boltzmann matrix is decomposed into two factors by introducing an intermediate index, denoted by the blue squares. $(c)$ The four matrices surrounding each vertex are contracted, with the original spins traced out, to form the initial tensor.}
    \label{fig:charactor_expansion}
\end{figure}

The character expansion introduces new indices by factorizing the Boltzmann matrix. For example, the Ising transfer matrix in Eq.~\eqref{eq:M-matrix} may be decomposed using an eigenvalue decomposition as:
\begin{align*}
    M &= U \Lambda U^\dagger
\end{align*}
where $\Lambda = \mathrm{diag}[2\cosh(\beta), 2\sinh(\beta)]$. If we now define $L=U\sqrt{\Lambda},\, R=\sqrt{\Lambda} U^\dagger$, then the Boltzmann matrix can be written as
$$M = LR.$$
with a new intermediate index inserted between $L$ and $R$ as illustrated in Fig.~\ref{fig:charactor_expansion} $(a$-$b)$. Once this is done, the original spin degrees of freedom can be traced out, leaving an initial tensor built from the four matrices surrounding each vertex. This basis has two clear advantages. First, it gives a discrete basis in irreducible representations even when the original spins themselves are continuous. Second, it makes the symmetry of the model manifest, which means one can use symmetric, block-diagonal tensor networks and enjoy a substantial boost in efficiency. Moreover, using a tensor network representations that is manifestly covariant with the symmetry of the underlying problem ensures that a whole class of perturbations stemming from numerical inaccuracies -- that are not allowed by the symmetry -- do not contribute errors to our final results. In the following, we elaborate on this construction through several examples. In particular, we review the Ising model, models with $\mathbb{Z}_q$ and other discrete symmetries, models with continuous fields and continuous group symmetries, Gauge Theories with Abelian and Non-abelian symmetries \cite{tensorfieldtheoryyannick} and Fermionic degrees of Freedom \cite{Gu:2010yh, akiyama2021more, mortier2025fermionic}.

The backbone of all calculations in TNRKit.jl is TensorKit.jl \cite{tensorkit}. It provides the low-level implementations for storing and manipulating symmetric tensors. For a thorough explanation on how to encode symmetric tensors in the framework of TensorKit, we refer the reader to the TensorKit \href{https://quantumkithub.github.io/TensorKit.jl/stable/}{documentation}\footnote{\href{https://quantumkithub.github.io/TensorKit.jl/stable/}{https://quantumkithub.github.io/TensorKit.jl/stable/}}, particularly the appendix called ``\textit{A symmetric tensor deep dive: constructing your first tensor map}''.

\subsubsection{Revisiting the Ising model}
\label{sec:revisitingtheisingmodel}
Let us now revisit the Ising model, this time building the initial tensor through the character expansion. For each edge $(x,\hat\mu)$, we perform a high-temperature, or strong-coupling, expansion\footnote{In this case the expansion is exact because the radius of convergence for the Taylor series of \(e^x\) is infinite, allowing us to refactor it in the way we did.} of the Boltzmann weight using the identity
\begin{equation*}
    e^{\beta \sigma \sigma'} = \cosh\beta\,\bigl(1+\sigma\sigma'\tanh\beta\bigr).
\end{equation*}
This introduces a binary link variable $n_{x,\hat\mu}\in\{0,1\}$, allowing us to write
\begin{equation*}
    e^{\beta \sigma_x \sigma_{x+\hat\mu}}
    =
    \cosh(\beta)
    \sum_{n_{x,\hat\mu}=0}^{1}
    \left(\sigma_x \sqrt{\tanh\beta}\right)^{n_{x,\hat\mu}}
    \left(\sigma_{x+\hat\mu} \sqrt{\tanh\beta}\right)^{n_{x,\hat\mu}}.
\end{equation*}
Inserting this expansion on every link, the partition function becomes
\begin{equation}
    Z = \cosh(\beta)^{2V}
    \sum_{\{\sigma=\pm 1\}}
    \sum_{\{n_{x,\hat\mu}=0,1\}}
    \prod_x
    \prod_{\mu=1}^{2}
    \left(\sigma_x
    \sqrt{\tanh\beta}\right)^{n_{x,\hat\mu} + n_{x-\hat\mu,\hat\mu}},
\end{equation}
where $V$ is the number of lattice sites, and the factor of $2$ in the exponent of $\cosh(\beta)$ reflects the two lattice directions. At each site $x$, the spin $\sigma_x$ appears in the four surrounding link factors. Collecting these contributions, the spin sum at site $x$ takes the form
\begin{equation}
    \sum_{\sigma=\pm 1}
    \prod_{\mu=1}^{2}
    \left(\sigma_x\sqrt{\tanh\beta}\right)^{n_{x,\hat\mu} + n_{x-\hat\mu,\hat\mu}}
    =
    \left(\tanh\beta\right)^{\frac{1}{2}\sum_\mu (n_{x,\hat\mu}+n_{x-\hat\mu,\hat\mu})}
    \sum_{\sigma=\pm 1}\sigma_x^{N_x},\label{eq:site_sum}
\end{equation}
where
\begin{equation*}
    N_x \equiv \sum_\mu \bigl(n_{x,\hat\mu}+n_{x-\hat\mu,\hat\mu}\bigr)
\end{equation*}
is the total occupation number of the four edges meeting at $x$. The remaining spin sum is easy to evaluate:
\begin{equation*}
    \sum_{\sigma=\pm 1}\sigma_x^{N_x}
    =
    2\,\delta_{N_x \bmod 2,\;0}.
\end{equation*}
It vanishes unless $N_x$ is even.

This is simply the lattice manifestation of the $\mathbb{Z}_2$ symmetry. In the dual link-variable language, it becomes the constraint
\begin{equation}
    n_{x,\hat 1}+n_{x-\hat 1,\hat 1}+n_{x,\hat 2}+n_{x-\hat 2,\hat 2}
    =
    0
    \pmod 2,\label{eq:z2_constraint}
\end{equation}
which must hold at every vertex. Equivalently, if we define the outgoing flux by
\begin{equation*}
    n_{x,\mathrm{out}} = n_{x,\hat 1}+n_{x,\hat 2}
\end{equation*}
and the incoming flux by
\begin{equation*}
    n_{x,\mathrm{in}} = n_{x-\hat 1,\hat 1}+n_{x-\hat 2,\hat 2},
\end{equation*}
then the constraint reads
\begin{equation*}
    n_{x,\mathrm{in}} \equiv n_{x,\mathrm{out}} \pmod 2.
\end{equation*}
In other words, the character expansion turns the original spin model into a theory of conserved $\mathbb{Z}_2$ flux living on the edges.
\begin{figure}[tb]
    \centering
    \includegraphics[width=\linewidth]{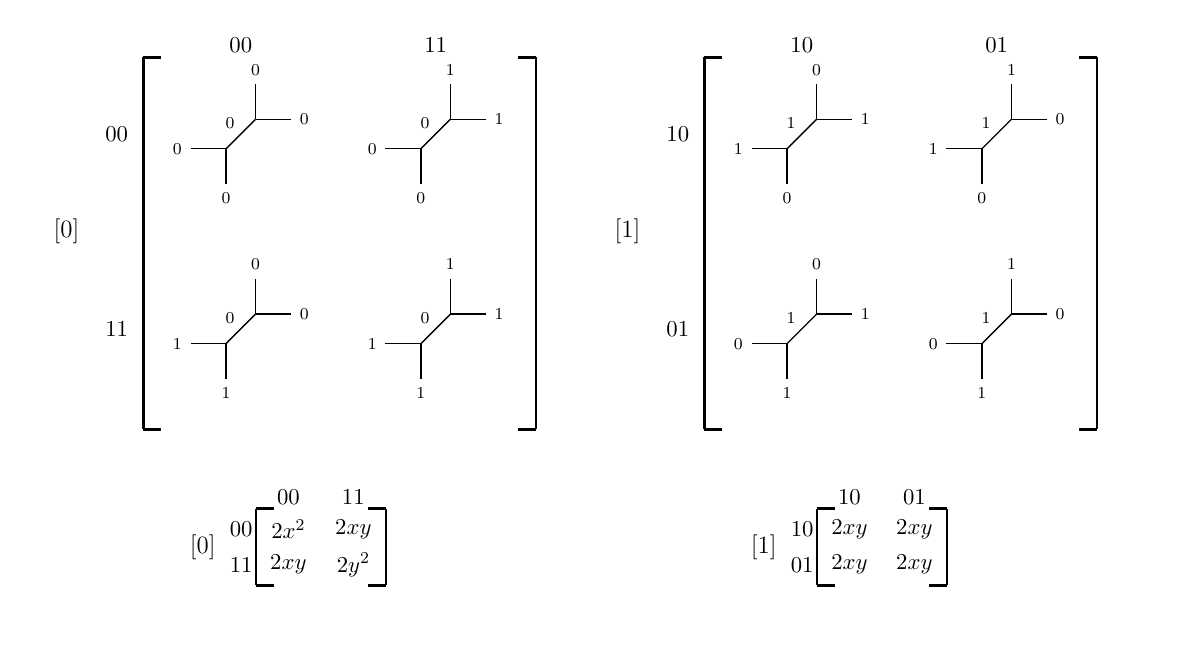}
   
    \caption{Block structure of the classical Ising tensor in the $\mathbb{Z}_2$-symmetric basis, with $x = \cosh\beta$ and $y = \sinh\beta$. \emph{Top:} each matrix entry shown as the corresponding fusion tree (leg index configuration). \emph{Bottom:} the same blocks written explicitly in terms of $x$ and $y$.}    
    \label{fig:z2symmetricising}
\end{figure}
Substituting Eqs.~\eqref{eq:site_sum}--\eqref{eq:z2_constraint} back into the partition
function, all spin degrees of freedom can be integrated out and we are just left with a sum over link variables. We define the rank-4 tensor at each site $x$~\cite{Liu_2013}, 
\begin{equation}
    T_{i\,j\,k\,l} =
    2\left(\tanh\beta\right)^{(i+j+k+l)/2}
    \,\delta_{(i+j+k+l)\bmod 2,\;0}
    \label{eq:init_isingtensor}
\end{equation}
where the indices $(i,j,k,l) = (n_{x-\hat{1},\hat{1}},\, n_{x-\hat{2},\hat{2}},\,
n_{x,\hat{2}},\, n_{x,\hat{1}}) \in \{0,1\}^4$ label the link variables on the \textbf{left,
bottom, top, and right bonds}\footnote{This ordering is the convention all 2d partition function tensors obey in TNRKit.} of site $x$, respectively. The partition function is then expressed as a
\emph{tensor network},
\begin{equation}
    \boxed{
    Z = 2^V \cosh(\beta)^{2V}
    \operatorname{tTr}\!\left[\bigotimes_x T^{(x)}\right]
    }
    \label{eq:Z_tensor}
\end{equation}
where $\operatorname{tTr}[\cdots]$ denotes a full contraction of all shared link indices between neighbouring tensors (i.e. a \emph{tensor trace}). The tensor $T^{(x)}$ is the same everywhere on an infinite square lattice. Because all indices are binary ($\chi = 2$), the tensor $T$ is a $2\!\times\!2\!\times\!
2\!\times\!2$ array with only eight non-zero entries, fixed entirely by the
$\mathbb{Z}_2$ constraint and the Boltzmann weight. 

In TNRKit, the initial tensor for the Ising model is defined in a block-structured form, as illustrated in Fig.~\ref{fig:z2symmetricising}. The tensor is block-diagonal in the intermediate index -- the coupled charge -- corresponding to
$n = n_{x,\mathrm{in}} = n_{x,\mathrm{out}}.$
The diagonal leg, or fusion channel, shown in Fig.~\ref{fig:z2symmetricising} represents this same index.\\

\noindent As an example we show the source code for the \(\mathbb{Z}_2\) symmetric Ising model initial tensor in TNRKit:

\noindent\includegraphics[page=3]{figures/TNRKit-paper-codesnippets.pdf}

\noindent TNRKit provides symmetric, and non-symmetric implementations for the classical Ising model's partition function.\\\\
\includegraphics[page=5]{figures/TNRKit-paper-codesnippets.pdf}

\subsubsection{Discrete \texorpdfstring{$\mathbb{Z}_q$}{Zq} Symmetry: Clock Models}
The $q$-state clock model is a natural generalisation of the Ising model ($q=2$) in which the spins at each site take one of $q$ equally spaced values on the unit circle, $\sigma_x = e^{2\pi i n_x / q}$ with $n_x \in \{0, 1, \ldots, q-1\}$.  The partition function on a square lattice is
\begin{equation}
    Z = \sum_{\{n_x\}} \prod_{(x,\hat\mu)}
        e^{\beta \cos\left(\frac{2\pi}{q}(n_x - n_{x+\hat\mu})\right)}
\end{equation}
To derive the tensor network for the clock model, we perform a \emph{$\mathbb{Z}_q$ discrete Fourier transformation} (a character expansion for finite $\mathbb{Z}_q$ groups) on each bond factor. Because the bond weight depends only on the difference $n_x - n_{x+\hat\mu} \bmod q$, it can be expanded in the irreducible characters of $\mathbb{Z}_q$, which are the $q$-th roots of unity $\chi_k(n) = e^{2\pi i k n/q}$ for $k \in \{0,\ldots,q-1\}$:

\begin{equation}
    e^{\beta \cos (2 \pi m/q)} = \sum_{k=0}^{q-1}  T_k \, e^{2\pi i k m/q}, 
    \qquad 
     T_k = \frac{1}{q}\sum_{m=0}^{q-1}
                 e^{\beta\cos(2\pi m/q)} e^{-2\pi i k m/q}
\end{equation}
where $m = n_x - n_{x + \hat\mu} \bmod q$ and $T_k$ are the Fourier (or character) coefficients. Inserting this at every bond, we can carry out the sum over spins at site $x$ which enforces a $\mathbb{Z}_q$ conservation law at every vertex; i.e. the total outgoing flux equals
the total incoming flux modulo $q$.
\begin{equation}
    \sum_{n_x = 0}^{q-1} e^{\frac{2\pi i}{q} n_x \left(k_{x,\hat 1} + k_{x,\hat 2}
       - k_{x-\hat 1,\hat 1} - k_{x-\hat 2, \hat 2}\right)} = q \, \delta_{k_{x,\hat 1} + k_{x,\hat 2}
       - k_{x-\hat 1,\hat 1} - k_{x-\hat 2, \hat 2} \; \pmod q, \;0}
\end{equation}
The rank-4 tensor at each site is the following, 
\begin{equation}
    T_{ijkl} = q\, T_i T_j  T_k T_l\,
               \delta_{i+j-k-l \pmod  q,\;0}
    \label{eq:clock_tensor}
\end{equation}
With each index running over $\{0,\ldots,q-1\}$ and a bond dimension of $\chi = q$, the $\mathbb{Z}_q$ symmetry is manifest in the Kronecker delta. For $q = 2$, this simply reduces to the Ising model's initial tensor derived in \ref{sec:Isingmodelinitialtensor}\\

\noindent TNRKit provides symmetric, and non-symmetric implementations for the classical clock model's partition function.\\\\
\includegraphics[page=6]{figures/TNRKit-paper-codesnippets.pdf}

\subsubsection{Continuous Fields: \texorpdfstring{$\phi^4$}{phi4} Theory}

For models with a continuous scalar field $\phi_x \in \mathbb{R}$, the strong-coupling expansion of the previous sections must be replaced by a different strategy. As a concrete example, we consider the two-dimensional $\phi^4$ theory with lattice action
\begin{equation}
    S[\phi] = \sum_x \left[
        \frac{m^2}{2}\phi_x^2 + \lambda \phi_x^4
        - \kappa \sum_{\hat\mu} \phi_x \phi_{x+\hat\mu}
    \right],
\end{equation}
where $\kappa$ controls the hopping strength and $m^2$, $\lambda$ are the mass and quartic coupling. The hopping term at each link is expanded in a suitable basis that decouples the two fields; this happens to be the Taylor expansion basis of polynomials. The path integral is as follows:
\begin{equation}
    Z = \int \prod_{x,\hat\mu} d \phi_x e^{-\frac{m^2}{2}\phi_x^2 - \lambda \phi_x^4} e^{\kappa \phi_x \phi_{x+\hat\mu}}.
\end{equation}
The hopping term in this path integral can be expanded as, 
\begin{equation}
    e^{\kappa \phi_x \phi_{x+\hat\mu}}
    = \sum_{n=0}^{N_{\rm cut}} \frac{\kappa^n}{n!} \phi_x^n \phi_{x+\hat\mu}^n,
    \label{eq:hop_expansion}
\end{equation}
where a link variable $n_{x,\hat\mu} \in \{0,\ldots,N_{\rm cut}\}$ is introduced for each
bond. After integrating out the local field $\phi_x$ using Eq.~\eqref{eq:hop_expansion}, the
resulting rank-4 tensor is
\begin{equation}
    T_{ijkl} = \int_{-\infty}^{\infty} d\phi\;
               e^{-m^2\phi^2/2 - \lambda\phi^4}\,
               \frac{\kappa^{i+j+k+l}}{\sqrt{i!\,j!\,k!\,l!}}\,
               \phi^{i+j+k+l},
    \label{eq:phi4_tensor}
\end{equation}

The Gaussian integral is only non-zero for even values of $i + j + k + l$, which translates the $\mathbb{Z}_2$ symmetry in the action to the tensor network. An alternative approach for continuous variables is the Gauss quadrature method \cite{kadoh2018tensor, kadoh2019tensor} (which is the same as the checkerboard method in spirit but for continuous variables). We prefer using a Taylor expansion over using the Gauss quadrature method because the $\mathbb{Z}_2$ symmetry is manifest in the initial tensor. Nevertheless, both options are available in TNRKit.\\
\noindent\includegraphics[page=7]{figures/TNRKit-paper-codesnippets.pdf}\\
The integer parameter $K$ controls the order of the Taylor expansion (\(\mathbb{Z}_2\)-symmetric implementation) or the number of Gauss–Quadrature points (non-symmetric implementation).

\paragraph{Character expansion and continuous group symmetries.}
When the action possesses a continuous group symmetry $G$, for instance $U(1)$, $SU(2)$, or $SU(N)$ in lattice gauge theories or nonlinear sigma models, the most systematic route to building a tensor network is using the \emph{character expansion}. The bond weight $e^{-S_{\rm link}(g)}$, which depends on a group element $g \in G$, is expanded in the complete basis of irreducible characters,
\begin{equation}
    e^{-S_{\rm link}(g)} = \sum_{R} c_R \chi_R(g),
    \qquad c_R = d_R \int_G dg\; e^{-S_{\rm link}(g)}\chi_R^*(g),
    \label{eq:char_expansion}
\end{equation}
Here $R$ labels irreducible representations (irreps), $d_R$ is the dimension of $R$,
$dg$ is the Haar measure, and $\chi_R(g) = \mathrm{tr}_R(g)$ is the character. For
$U(1)$ the irreps are labelled by integers $n \in \mathbb{Z}$ and the expansion reduces
to a standard Fourier series; for $SU(2)$ they are labelled by half-integer spins
$j \in \{0, \frac{1}{2}, 1, \ldots\}$. 

After inserting Eq.~\eqref{eq:char_expansion} on every bond, a link variable $R_{x,\hat\mu}$ (the irrep label) is introduced on each bond. The group integration at each vertex, using the Peter-Weyl theorem, enforces a constraint such that the tensor product of the incoming irreps contains a singlet (trivial representation). This is the non-abelian generalisation of the $\mathbb{Z}_q$ conservation law. Crucially, the irrep label $R$ plays the role of the index on each tensor leg such that the bond dimension is $\chi = \sum_R d_R^2$ where $R \leq R_{\rm max}$ is a truncation in the representation space. TNRKit builds on TensorKit's symmetry-aware tensor arithmetic to enforce and preserve the block-diagonal structure in representation space across all steps of a TNR algorithm, yielding significant reductions in memory and computational cost.

\subsubsection{Fermionic Models: Grassmann Tensor Networks}

Fermionic degrees of freedom require special treatment because the fields anticommute, \\
$\{\psi(n), \bar\psi(m)\} = \delta_{m,n}$. We briefly outline how fermionic path
integrals can be cast into a tensor network form using Grassmann variables. Consider a
general fermionic action in $d$ dimensions,
\begin{equation}
    S[\bar\psi, \psi] = \sum_{n \in \Lambda} \left[
        - \kappa \sum_{\nu=1}^d \left(
            \bar\psi(n)\,\psi(n+\hat\nu) + \bar\psi(n+\hat\nu)\,\psi(n)
        \right) + W[\bar\psi(n), \psi(n)]
    \right],
\end{equation}
where $W[\bar\psi, \psi]$ encodes all on-site terms such as the mass and quartic
interactions, and $\bar\psi$, $\psi$ are taken to be single-component fermionic fields
for simplicity. The partition function is
\begin{equation}
    Z = \left(\prod_{n \in \Lambda} \int d\bar\psi(n)\, d\psi(n)\right) e^{-S[\bar\psi,\psi]}.
\end{equation}
 
To decouple the hopping terms and isolate the degrees of freedom at each site, we
introduce auxiliary Grassmann fields $\eta_\nu(n)$, $\bar\eta_\nu(n)$ and $\zeta_\nu(n)$, $\bar\zeta_\nu(n)$ living on the bonds of the lattice. The hopping factors are then decomposed via Grassmann Gaussian integrals,
\begin{equation}
    \begin{aligned}
        e^{\kappa\,\bar\psi(n)\psi(n+\hat\nu)}
        &= \int d\bar\eta_\nu(n)\,d\eta_\nu(n)\;
           e^{-\bar\eta_\nu(n)\eta_\nu(n)}\,
           e^{\sqrt{\kappa}\,\bar\psi(n)\eta_\nu(n)}\,
           e^{-\sqrt{\kappa}\,\psi(n+\hat\nu)\bar\eta_\nu(n)},
        \\[4pt]
        e^{\kappa\,\bar\psi(n+\hat\nu)\psi(n)}
        &= \int d\bar\zeta_\nu(n)\,d\zeta_\nu(n)\;
           e^{-\bar\zeta_\nu(n)\zeta_\nu(n)}\,
           e^{-\sqrt{\kappa}\,\bar\psi(n+\hat\nu)\bar\zeta_\nu(n)}\,
           e^{-\sqrt{\kappa}\,\psi(n)\zeta_\nu(n)}.
    \end{aligned}
\end{equation}
With this decomposition, the physical fields $\bar\psi(n)$ and $\psi(n)$ appear only in on-site factors and can be integrated out independently at each vertex, leaving a local Grassmann tensor
\begin{equation}
    \mathcal{T}(n) = \int d\bar\psi\, d\psi\; e^{-W[\bar\psi,\psi]}
    \prod_{\nu=1}^d
    e^{\sqrt{\kappa}\,\bar\psi\,\eta_\nu(n)}\,
    e^{-\sqrt{\kappa}\,\psi\,\zeta_\nu(n)}\,
    e^{-\sqrt{\kappa}\,\bar\psi\,\bar\zeta_\nu(n-\hat\nu)}\,
    e^{-\sqrt{\kappa}\,\psi\,\bar\eta_\nu(n-\hat\nu)},
\end{equation}

whose legs are the bond Grassmann variables $\eta_\nu$, $\zeta_\nu$ and their conjugates.
The partition function is then expressed as a \emph{Grassmann tensor network},
\begin{equation}
    Z = \operatorname{gTr}\!\left[
        \prod_{n\in\Lambda}
        \mathcal{T}_{\Psi_1(n)\cdots\Psi_d(n)\,
                     \bar\Psi_d(n-\hat{d})\cdots\bar\Psi_1(n-\hat{1})}
    \right],
\end{equation}
where $\Psi_\nu = (\eta_\nu, \zeta_\nu)$ and $\bar\Psi_\nu = (\bar\eta_\nu,\bar\zeta_\nu)$ are the composite Grassmann indices on each bond. The Grassmann trace, which correctly accounts for fermionic exchange statistics, is defined as
\begin{equation}
    \operatorname{gTr}[\cdots] =
    \prod_{n\in\Lambda}\prod_{\nu=1}^d
    \int d\bar\Psi_\nu(n)\, d\Psi_\nu(n)\;
    e^{-\bar\Psi_\nu(n)\Psi_\nu(n)}\;(\cdots).
\end{equation}
 
A Grassmann tensor is non-vanishing only when it is Grassmann-even, meaning its entries
carry a fermionic $\mathbb{Z}_2$ grading that forces a block-diagonal structure in the
space of occupied and unoccupied fermionic modes — precisely analogous to the $\mathbb{Z}_2$ block structure encountered in the Ising tensor in Eq.~\ref{eq:z2_constraint}. TensorKit represents this structure natively through fermionic $\mathbb{Z}_2$ graded vector spaces, so that each tensor index is spanned by

\noindent \includegraphics[page=4]{figures/TNRKit-paper-codesnippets.pdf}

\section{From the Renormalization Group to Tensor Networks}
\label{sec:tnr}
\begin{figure}[tb]
    \centering
    \includegraphics[page=14,width=0.8\linewidth]{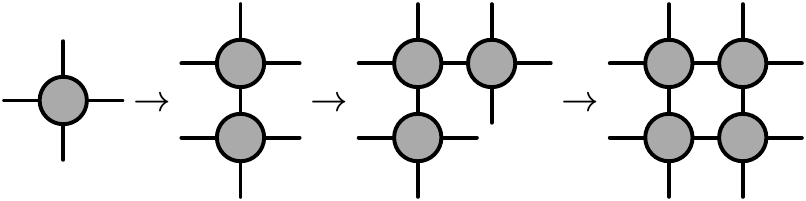}
    \caption{Illustration of a two-dimensional tensor network contraction. As more tensors get contracted, the computational and memory costs explodes.}
    \label{fig:intro_2dcontraction}
\end{figure}

After learning how to encode your favorite partition functions on a computer, the next question is obvious: how do you actually evaluate them? Unfortunately, there is a no-go theorem that rules out their exact evaluation. This is one of the most fundamental obstacles, appearing both in the quantum and classical settings. Figure \ref{fig:intro_2dcontraction} shows a typical example encountered in tensor-network contractions. As more tensors are joined together, the intermediate tensors grow to higher and higher rank, until one is left with an enormous tensor that simply will not fit in your computer’s memory. Mathematically, contracting tensor networks is known to be \#P-complete~\cite {pcomplete_2007}.

In this context, the key task is to find approximation schemes that are both accurate and practical. Tensor renormalization group (TRG)~\cite{Levin_trg,HOTRG,BTRG,ATRG,TRG_lan_2019,PTMRG,triad_trg,Boundary_TRG_2019,fermion_TRG_2024,Gu:2010yh} and tensor network renormalization (TNR)~\cite{Gu_2009,Looptnr_2017,Hauru_2016,TNR_evenbly_2015,NNR_2024,tnrplus_2017,thermal&globaltnr,Evenbly:2007hxg,GILT} methods are such schemes, built upon the idea of RG.

\subsection{Levin-Nave TRG}

\begin{figure}[tb]
    \centering
    \includegraphics[page = 1, width=\linewidth]{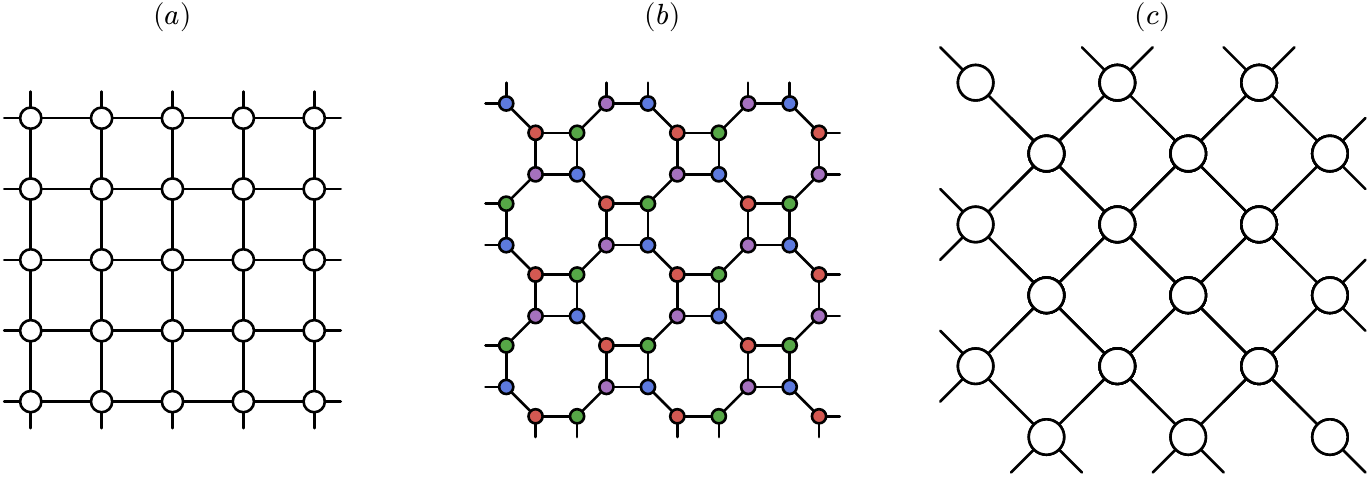}
    \caption{Levin–Nave TRG. Each four-leg tensor on the square lattice is split into two three-leg tensors, which are then recombined in a different pattern. The resulting tensor network contains half as many tensors as the original one.}
    \label{fig:Levin_TRG}
\end{figure}

TRG was born only about twenty years ago, making it a relatively young idea. White’s DMRG, by contrast, had already been around for more than a decade.\footnote{We should also mention the invention of the Corner Transfer Matrix Renormalization Group (CTMRG), introduced by Nishino and Okunishi in the mid-1990s~\cite{Nishino_1995,Nishino_1996,Nishino_1997}. CTMRG was one of the pioneering methods for contracting two-dimensional tensor networks efficiently. It remains widely used today, particularly in the contraction of two-dimensional tensor-network states such as PEPS~\cite{peps,Orus_2009}. Although this review focuses on a different branch of the story, CTMRG algorithms are also available in TNRKit.} Levin and Nave looked at the square-lattice tensor network in Fig.~\ref{fig:Levin_TRG}(a) and asked the obvious RG question: how do we coarse-grain this object? After all, the essence of the renormalization group is to reduce the number of degrees of freedom, while discarding the irrelevant information along the way. Their insight was to bring the singular value decomposition (SVD) into the game.\footnote{To be politically more correct, one should note that SVD had long been used in the DMRG community by this point. The contributions of Levin and Nave were to the application of these ideas to classical stat-mech models, where the notion of entanglement is far less obvious.} In quantum information, the SVD is a powerful diagnostic of entanglement: the singular values tell us which degrees of freedom matter most. In tensor networks, that makes it an ideal tool for separating the important information from the disposable clutter.

\begin{figure}[hb]
    \centering
    \includegraphics[page=15]{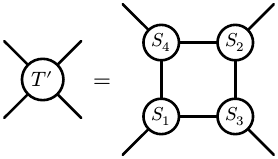}
    \caption{Combination step of the Levin-Nave TRG algorithm.}
    \label{fig:TRG-combination}
\end{figure}

The first step of the Levin-Nave TRG algorithm is a decomposition: each four-leg tensor is split into a pair of three-leg tensors via an SVD, as in Fig.~\ref{fig:Levin_TRG}(b) and Eq.~\eqref{eq:TRG-svd}. The two resulting tensors share a new bond, which carries the newly introduced ``entanglement" degrees of freedom. The singular values on this bond tell us which of these degrees of freedom matter most, and truncating to the largest $\chi$ of them is exactly the renormalization-group step: the unimportant information is discarded, while the important part is kept. In the second step, four of these three-leg tensors are combined into a new four-leg tensor, as shown in Fig.~\ref{fig:Levin_TRG}(c) and~\ref{fig:TRG-combination}. The result is a new tensor network, now expressed entirely in terms of these coarse-grained, entanglement-weighted degrees of freedom. Since each coarse-graining step reduces the number of tensors by a factor of two, after sufficiently many iterations, one is left with a single tensor. Contracting a single tensor is, of course, no longer \#P-complete. Problem solved.

This is the Levin--Nave TRG algorithm: the first of its kind, and still among the fastest TRG methods available. With a humble MacBook Air, it can calculate the free energy of the critical 2D classical Ising model to six digits of accuracy in about 0.2 seconds.

There is a good reason why this algorithm works so well. When using the Frobenius norm, the optimal approximation of a given tensor by another lower-rank tensor is given by SVD. In other words, under the condition that the rank of the target tensor \(T^\prime\)\footnote{the rank of \(T^{\prime}\) when viewing \(T^{\prime}\) as a matrix when we partition its indices by a cut along one of the diagonals.} is $\chi$, the singular value decomposition minimizes the following Frobenius norm:\footnote{This is known as the Eckart–Young–Mirsky theorem.}

\begin{align}
    \raisebox{-0.5\height}{\includegraphics[width=50mm, page=10]{figures/TNRKit-paper-diagrams.pdf}},
    \label{eq:svd_norm}
\end{align} 
through the truncation of the intermediate singular values as

\begin{equation}
    \raisebox{-0.5\height}{\includegraphics[page=11]{figures/TNRKit-paper-diagrams.pdf}}.
    \label{eq:TRG-svd}
\end{equation}
Note that we absorb the truncated singular values symmetrically in both new tensors:
\begin{align*}
    T &= U\cdot S \cdot V^{\dagger}\\
      &\approx U \cdot \tilde{S} \cdot V^{\dagger}\\
      &=(U\cdot \sqrt{\tilde{S}}) \cdot (\sqrt{\tilde{S}} \cdot V^{\dagger}) = S_1 \cdot S_2.
\end{align*}
The passage from $S$ to $\tilde{S}$ indicates the truncation: only the $\chi$ largest singular values of the diagonal matrix $S$ are retained, while the rest are discarded. The computational cost of the Levin-Nave TRG algorithm scales as \(\mathcal{O}(\chi^6)\).

Let us pause for a brief practical remark on how the partition function is actually computed. After \(n\) RG steps, the renormalized tensor represents an effective local tensor for a system of linear size \(L=(\sqrt{2})^n\). Tracing over the horizontal and vertical pairs of legs then gives the partition function on an \(L\times L\) torus, which we denote by \(Z(L,L)\).

\begin{figure}
    \centering
    \includegraphics[page=12]{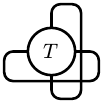}
    \caption{Traced partition function tensor used to normalise the tensor and calculate observables. When working with tensors in vector spaces that have non-trivial braiding rules, one should resolve the braiding explicitly.}
    \label{fig:tracenorm}
\end{figure}

In practice, one does not work directly with these tensors, since at every step they are normalized by the normalization factor $g^{(n)}=\frac{Z(L,L)}{Z(L/b,L/b)^{b^2}}$, with $b=\sqrt{2}$, (c.f. Figure \ref{fig:tracenorm}).
Combining these factors over successive RG steps, one reconstructs the partition-function density via \cite{Gu_2009}
\begin{align}
\frac{\ln Z(b^n,b^n)}{b^{2n}}
= \sum_{j=1}^n \frac{\ln g^{(j)}}{b^{2j}}.
\end{align}

\noindent When running a TRG or TNR scheme in TNRKit, by default these normalization factors get returned. You can use the \texttt{free\_energy} function to perform this sum for you.\\\\
\noindent\includegraphics[page=1]{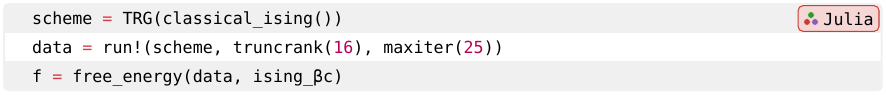}

\subsection{HOTRG}
In 2012, Xie. et. al \cite{HOTRG}. generalised the idea of Levin and Nave's TRG algorithm by using the higher-order singular value decomposition \cite{delathauwer2000}. Instead of splitting each tensor in the network into two and then combining four parts, they merge two neighbouring tensors into one. First in one lattice direction, and then in the other.
This algorithm has the advantage that it produces more accurate results for the free energy density (c.f. Section \ref{sec:benchmarks}), but at an increased computational cost of \(\mathcal{O}(\chi^7)\). Additionally, the ``HOTRG" algorithm can be used in any dimension \(d\), with the cost scaling as \(\mathcal{O}(\chi^{4d - 1})\).

\begin{figure}[tb]
    \centering
    \includegraphics[page=5]{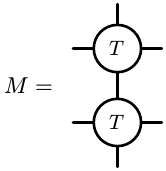}
    \caption{The HOTRG algorithm combines two tensors \(T\) into one tensor \(M\).}
    \label{fig:HOTRG-M}
\end{figure}

\begin{figure}[tb]
    \centering
    \begin{subfigure}{0.4\textwidth}
        \centering
        \includegraphics[scale=0.8, page=1]{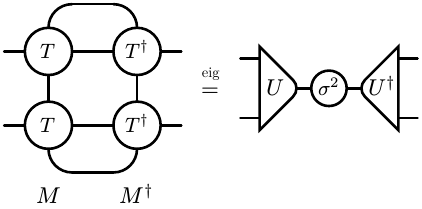}
        \caption{First isometry}
    \end{subfigure}
    \begin{subfigure}{0.4\textwidth}
        \centering
        \includegraphics[scale=0.8, page=2]{figures/TNRKit-paper-diagrams.pdf}
        \caption{Second isometry}
    \end{subfigure}
        \caption{Diagrams explaining the two ways to generate coarse-graining isometries for the HOTRG algorithm.}
    \label{fig:HOTRG-isometries}
\end{figure}

During coarse graining, instead of performing an SVD on one tensor \(T\), the HOTRG algorithm combines two of them into a single tensor \(M\) (c.f. Figure \ref{fig:HOTRG-M}). The next step is to multiply these 2 tensors with an isometry from both sides, to combine their legs in a (locally) optimal way. To get the left and right (or top and bottom) truncated singular vectors, which make up these isometries, one could use an SVD (\(M = U\sigma V^{\dagger}\)). Instead, we can use a truncated hermitian eigenvalue decomposition on \(M^{\dagger}M\) and \(MM^{\dagger}\), which is much more cost-effective (c.f. Figure \ref{fig:HOTRG-isometries}).

\begin{figure}[tb]
    \centering
    \begin{subfigure}{0.4\textwidth}
        \centering
        \includegraphics[scale=0.8,page=3]{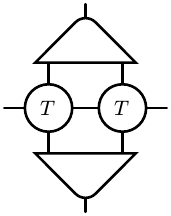}
        \caption{Horizontal coarse graining step}
    \end{subfigure}
    \begin{subfigure}{0.4\textwidth}
        \centering
        \includegraphics[scale=0.8,page=4]{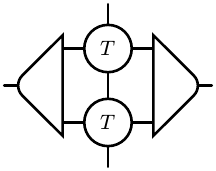}
        \caption{Vertical coarse graining step}
    \end{subfigure}
        \caption{Coarse-graining steps in the HOTRG algorithm}
    \label{fig:HOTRG-coarsegrain}
\end{figure}

We can then combine the two tensors \(T\) into one by applying the unitary that makes the smallest error (as defined by the sum of the discarded eigenvalues) on both sides of the pair (c.f. Figure \ref{fig:HOTRG-coarsegrain}).

One step of the HOTRG algorithm combines a vertical and horizontal step, effectively rescaling both lattice directions by 2, leading to a network that holds a quarter of the amount of tensors.

\noindent You can use the HOTRG algorithm to contract your own tensor network using the following code:\\
\includegraphics[page=2]{figures/TNRKit-paper-codesnippets.pdf}

\subsection{Other TRG algorithms}
There is a plethora of other TRG algorithms out there. Bond-weighted TRG \cite{BTRG} (\texttt{BTRG} in TNRKit) retains some singular values from the Levin–Nave TRG on the network bonds; it matches the computational cost of standard TRG while offering substantially improved accuracy, making it the perfect candidate for quick and accurate results. Anisotropic TRG \cite{ATRG} (called \texttt{ATRG} in TNRKit) is a particularly cheap algorithm that can be used in any dimension \(d\) as its cost scales as \(\mathcal{O}(\chi^{2d+1})\) but its results are rather inaccurate and should therefore be avoided in 2D. A more recent contribution is the Periodic Transfer Matrix Renormalization Group \cite{PTMRG}(PTMRG), which grows the system size linearly rather than exponentially, resulting in lower computational cost and a denser set of data points. Moreover, these algorithms can also be used for simulating path integrals in higher dimensions such as in 3D \cite{thermal&globaltnr, 3dzngaugetheory} and even in four dimensions \cite{triadrepatrg, samberger2025tensor, sugimoto4dqcd}. 

\paragraph{Limitation of TRG}
Despite its many successes, TRG is not without shortcomings. A major problem, already noted in Refs.~\cite{Levin_trg, Gu_2009}, is that it fails to reproduce the correct fixed points of ordered phases. Levin traced this back to the very nature of SVD-based TRG. A class of unphysical fixed-point tensors known as corner double line (CDL) tensors is known to be an exact fixed point of the TRG algorithm. Their structure is illustrated below:
\begin{align}
    \raisebox{-0.5\height}{\includegraphics[page=6]{figures/TNRKit-paper-diagrams.pdf}}\label{CDL_figure}
\end{align}
\begin{figure}[tb]
\centering
\includegraphics[page = 20, width=0.8\linewidth]{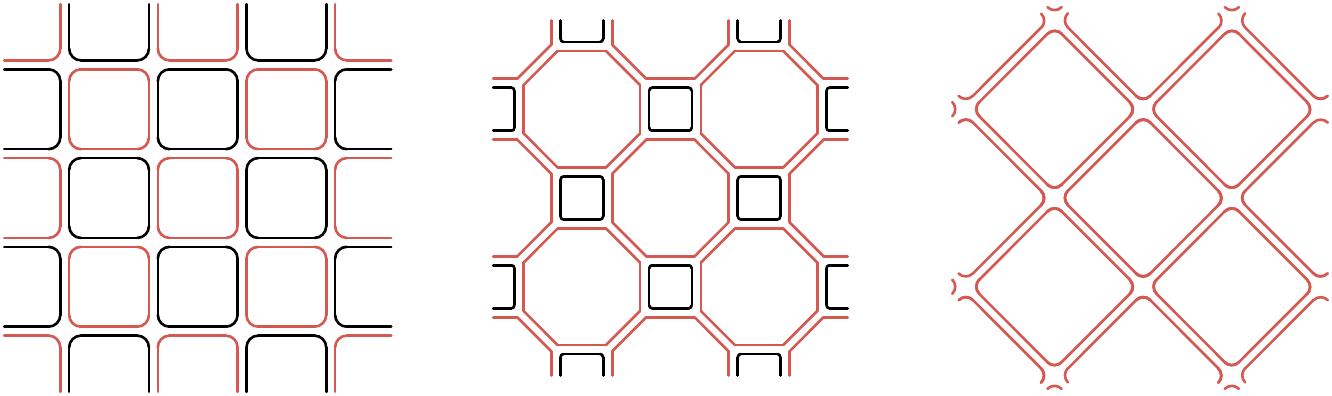}
\caption{The CDL tensors after TRG. The local loop, marked by a red square, persists after the TRG steps, indicating that ultraviolet information remains even after extensive coarse-graining.}
\label{CDL_tensors}
\end{figure}
Any local information that contains a component resembling a CDL tensor is not removed by TRG, as indicated by the red squares in Fig.~\ref{CDL_tensors}.
As the figure makes clear, ultraviolet information survives through successive coarse-graining steps instead of being washed away. These spurious degrees of freedom then consume valuable bond dimension, leaving fewer resources for the physics that actually matter. The result is a progressively poorer approximation after many RG steps. 

The CDLs encountered in two-dimensional tensor network simulations are actually a part of a broader phenomenon encountered in cyclic networks called internal correlations. A measure for these internal correlations is cycle or loop entropy, which is invariant under a choice of gauge \cite{gaugefixingcanonicalformsFET}. Removal of these internal correlations is perhaps the central issue in improving tensor network algorithms. Various algorithms, such as entanglement filtering \cite{Gu_2009} and full environment truncation \cite{gaugefixingcanonicalformsFET} have been proposed to remedy this.  

This limitation of TRG points to the need for more sophisticated coarse-graining schemes, going beyond the purely local tensor truncation of Eq.~\eqref{eq:svd_norm}. Such algorithms fall under the broader heading of tensor network renormalization, or TNR, whose purpose is precisely to remove this unwanted short-distance structure and thereby improve the accuracy of the coarse-graining procedure.

\subsection{Tensor Network Renormalization and LoopTNR}
What sets TNR methods apart from TRG methods comes down to two things: they coarse-grain the lattice while actively removing spurious local entanglement structures such as CDL tensors, and they use a larger unit cell in the tensor optimization.

The first attempt at this was Gu and Wen's Tensor Entanglement Filtering Renormalization (TEFR) algorithm in 2009, followed by Evenbly and Vidal's Tensor Network Renormalization papers starting in 2014. The current gold standard, however, is the LoopTNR algorithm published by Yang et al. in 2017. Beyond producing accurate observables, its real strength lies in the stability of the CFT spectrum it yields throughout coarse-graining -- it manages to remain at the unstable critical fixed point for a remarkably large number of coarse-graining steps.

Because the LoopTNR algorithm is still the state-of-the-art TNR algorithm, it is the subject of the remainder of this section, serving as an illustrative example for a TNR method.

The cost function of LoopTNR is:
\begin{equation}
     \raisebox{-0.475\height}{\includegraphics[page=7, scale=0.65]{figures/TNRKit-paper-diagrams.pdf}
    \label{eq:tnr_norm}}
\end{equation}
under the constraint that the diagonal legs of the new tensors $\{S_i\}, ~i=1,\dots8$ are of dimension \(\chi\).
In contrast to the SVD-based approach described in Eq.~\eqref{eq:svd_norm}, LoopTNR focuses on a two-by-two unit cell which can be composed of two distinct tensors \(T_A\) and \(T_B\). Using this cost function, an 8-leg tensor (as shown on the left side) is approximated through the contraction of eight 3-leg tensors, denoted as $\{S_1, \dots, S_8\}$ (as depicted on the right side).

As an initial guess for the new tensors, we use a simple SVD as in the TRG algorithm. Let the left and right sides of Eq.~\eqref{eq:tnr_norm} be $|\Psi_A\rangle$, consisting of a two-by-two patch of the original network, and $|\Psi_B\rangle$, consisting of the new -- to be optimized -- tensors. Then, the cost function is equal to:
\begin{align}
\label{eq:looptnr_cost}
&\mathcal{L}(\{S_i\}) = \left\| |\Psi_A\rangle-|\Psi_B\rangle \right\|^2_F
=\langle\Psi_A|\Psi_A\rangle + \langle\Psi_B|\Psi_B\rangle-\langle\Psi_A|\Psi_B\rangle-\langle\Psi_B|\Psi_A\rangle,
\end{align}
Now, we optimize the new tensors to minimize this cost function. For convenience, we define $C$, $N_i$, $W_i$, and $W_i^\dag$ as follows:
\begin{align*}
    \includegraphics[page=8, scale=0.6]{figures/TNRKit-paper-diagrams.pdf}.
\end{align*}
This allows to rewrite Eq.~\eqref{eq:looptnr_cost} as 
\begin{align}
\mathcal{L}(\{S_i\})&=\||\Psi_A\rangle-|\Psi_B\rangle\|^2\nonumber\\
&= C + (S^j)^\dag N_j S^j - W^\dag_jS^j - (S^j)^\dag W_j.
\end{align}
for any \(j \in \{1, \dots, 8\}\).
This function is quadratic in the new tensors. If we fix every tensor except for $S^j$, the minimum can be found by solving $\frac{\partial \mathcal{L}}{\partial {S^j}^{\dagger}}=0$. The minimum of the cost function where we keep all tensors except for \(S^j\) fixed can therefore be found by solving the linear problem\footnote{In practice, we find that using a dense linear solver to solve the problem is faster than using the usual Krylov methods. TNRKit provides the user with the option to use both.}:
\begin{align}
    N_j S_j = W_j.
\end{align}
We then iterate over the different \(S_i\)'s, solving the linear problem for each one until we have reached convergence (or a predetermined maximum number of iterations).\\

\begin{figure}
    \centering
    \centering
    \begin{subfigure}{0.48\textwidth}
    \centering
        \includegraphics[width=0.8\linewidth, page=21]{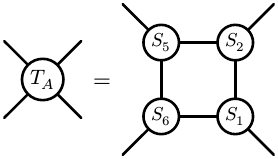}
        \caption{Recombination step for \(T_A\)}
    \end{subfigure}
    \hfill
    \begin{subfigure}{0.48\textwidth}
        \centering
        \includegraphics[width=0.8\linewidth, page=22]{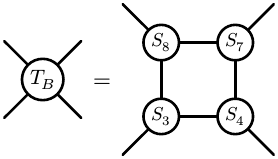}
        \caption{Recombination step for \(T_B\)}
    \end{subfigure}
        \caption{LoopTNR recombination steps.}
    \label{fig:LoopTNRrecombination}
\end{figure}

To perform the coarse-graining step we combine the \(S_i\)'s in two different ways, giving rise to the the updated \(T_A\) and \(T_B\) tensors as shown in Figure~\ref{fig:LoopTNRrecombination}.

There are a number of computational tricks one can employ when implementing the LoopTNR algorithm, which can be found in Ref.~\citenum{bao2019}. Luckily, these tricks are already implemented in TNRKit.jl.

To combat spurious local correlations, the LoopTNR algorithm implements an \textit{Entanglement Filtering} (EF) algorithm. When the tensors have a CDL structure as shown in Eq.~\eqref{CDL_figure}, \(|
\Psi_A\rangle\) can be expressed as:
\begin{align*}
    \raisebox{-0.475\height}{\includegraphics[scale=0.6, page=9]{figures/TNRKit-paper-diagrams.pdf}},
\end{align*}
where the red loop corresponds to the local loop in Fig.~\ref{CDL_tensors}. In scenarios where the red loop encompasses $n$ dimensions, each tensor within this red loop is associated with a rank-$n$ diagonal matrix:
\begin{align}
    \raisebox{0.2\height}{\includegraphics[page=16]{figures/TNRKit-paper-diagrams.pdf}} \quad= \quad \begin{pmatrix}
        \lambda_1& 0& \dots& 0\\
        0& \lambda_2& \dots& \dots\\
        \dots & \dots & \dots & 0 \\
        0 & \dots & 0 & \lambda_n
    \end{pmatrix}.\label{red_part}
\end{align}
where $\lambda_i > \lambda_{i+1} ~ \forall i$. The primary objective in entanglement filtering is to effectively compress this matrix to a rank-1 configuration. This compression is achieved by constructing a projector between $T_i$ and $T_{i+1}$ (where we identify \(T_1\) with the top left \(T_A\) of Eq.~\eqref{eq:tnr_norm}, \(T_2\) with the top right \(T_B\), \(T_3 \equiv T_A\) and \(T_4 \equiv T_B\) in a clockwise fashion) that targets the subspace corresponding to $\lambda_1$, the largest singular value. To do this, we use a QR decomposition. 
Consider the procedure of inserting a projector between tensors 
$T_4$ and $T_1$ in a TNR setup, where we denote $T_{i+4}=T_i$ for cyclic consistency. The first step involves placing a rank-$d$ identity matrix, denoted as $L_1^{[1]}$
 , to the left of $T_1$. Subsequently, we apply a QR decomposition to the tensor product of $L_1^{[1]}$ and $T_1$, resulting in:
\begin{align}
L_1^{[1]} T_1 = \tilde{T}_1 L_1^{[2]},
\end{align}
where $\tilde{T}_1 $ is an orthogonal matrix and $L_1^{[2]}$ is an upper triangular matrix.
The next step involves normalizing $L_1^{[2]}$ and repeating a similar QR decomposition process with $L_1^{[2]}$ and $T_2$, and then proceeding with 
$L_1^{[3]}$ and $T_3$. This iterative process is continued until convergence is achieved, resulting in the final projector 
$L_1^{[\infty]}$ (The convergence is checked when $L_1$ comes back to between $T_4$ and $T_1$.
During this process, $L$ accumulates the matrix in Eq.~\eqref{red_part} to end up having 

\begin{align*}
\lim_{m\rightarrow\infty}
\begin{pmatrix}
\lambda_1^m & 0 &\cdots&0 \\
0 & \lambda_2^m&\cdots&\cdots \\
\cdots&\cdots &\cdots&0\\
0&\cdots &0&\lambda_n^m
\end{pmatrix}
\propto 
\begin{pmatrix}
1 & 0 &\cdots&0 \\
0 & 0&\cdots&\cdots \\
\cdots&\cdots &\cdots&0\\
0&\cdots &0&0
\end{pmatrix}.
\end{align*}
We repeat the same thing to the left starting from $R_4^{[1]}$ and $T_4$ to obtain $R_4^{[\infty]}$. Finally, we obtain the projectors using SVD as follows~\cite{wang2011cluster,projector_corboz,Boundary_TRG_2019}:
\begin{align}
    L_1^{[\infty]}R_4^{[\infty]} = U_{41}\Lambda_{41}V^\dag_{41}\nonumber,\\
    P_{4R} = R^{\infty}_4V_{41}\frac{1}{\sqrt{\Lambda_{41}}},\\
    P_{1L} = \frac{1}{\sqrt{\Lambda_{41}}} U^\dag_{41}L_1^{\infty},
\end{align}
Having obtained all projectors, we redefine $T_{A/B}$ by contracting them with four projectors as:

\begin{align*}
    \raisebox{-0.5\height}{\includegraphics[page=13, width=60mm]{figures/TNRKit-paper-diagrams.pdf}}.
\end{align*}

This procedure reduces the CDL loop structure, allowing the LoopTNR algorithm to obtain stable CFT data for many TNR iterations. 
For more details, consult the original paper, Ref.~\citenum{Looptnr_2017}.

A recent paper~\cite{NNR_2024} by Homma \emph{et al.} takes a further step towards removing loop correlations by adding a nuclear-norm term for the $S_i$ tensors to the cost function. The resulting optimization problem can be handled with the alternating direction method of multipliers (ADMM). This scheme is also implemented in TNRKit.jl, and in practice it yields an even more stable spectrum, allowing higher scaling dimensions to be resolved (c.f. Section~\ref{sec:benchmarks}).

To apply LoopTNR to your own tensor network in TNRKit.jl, you may use the following code:

\noindent\includegraphics[page=9]{figures/TNRKit-paper-codesnippets.pdf}

\section{Extracting CFT spectrum from fixed-point tensors}\label{sec:cft}
A fixed-point tensor represents the physics in the macroscopic limit. Therefore, the data of a phase may be completely encoded in such a tensor. In the pioneering work of Ref.~\citenum{Gu_2009}, techniques were introduced to extract the ground-state degeneracy and scaling dimensions from a fixed-point tensor. These techniques were later significantly improved upon to extract more data of the CFT, such as conformal spins \cite{bao2019,Hauru_2016} and even structure constants \cite{Li_2022,ueda2023fixedpointtensorfourpointfunction} with high precision. Understanding the structure of the fixed-point tensor itself is also an important problem; see for example Refs.~\citenum{ueda2023fixedpointtensorfourpointfunction, Cheng:2023kxh, bao2025tensorcomplexrenormalizationgeneralized}. Moreover, recent years have seen significant progress in using tensor networks as a framework for putting the renormalization group itself on firmer footing~\cite{slava1,slava2,slava3,slava4,slava5,Ebel1}. In this setting, TNR provides a natural way to derive upper bounds on the growth of perturbations with scale around a fixed-point tensor.

In this section, we review how to extract scaling dimensions, central charges, and conformal spins. More importantly, we will introduce a geometric viewpoint of local tensors, such that different choices of the transfer matrix can be understood in a unified perspective. From the geometric perspective, we introduced a method, called the jigsaw trick, to systematically calculate higher scaling dimensions and conformal spins from a fixed-point tensor. These methods have been implemented in TNRKit.

\subsection{Geometric understanding of the fixed-point tensor}
At the RG fixed point, a single local tensor

$$ T_*: V_l \otimes V_d \leftarrow V_u\otimes V_r$$
approximates -- up to controllable errors -- the contraction of a large block of the initial tensor network~\cite{Levin_trg}:
\[
\diagram[2]{
    \draw (-1,0) node[left]{$V_l$} -- (1,0) node[right]{$V_r$};
    \draw (0,-1) node[below]{$V_d$} -- (0,1) node[above]{$V_u$};
    \filldraw[fill=white, draw = black] (-.2,-.2) rectangle node{$T_*$} (.3,.3);
}
\approx 
\diagram[2]{
    \foreach \i in {-0.9, -0.8, ..., 0.9}
    {
    \draw[black!20] (-1,\i) -- (1,\i);
    }
    \foreach \j in {-0.9, -0.8, ..., 0.9}
    {
        \draw[black!20] (\j,-1) -- (\j,1);
    }
    \node[above] at (0,1) {$W^1_{u}\otimes W^2_{u} \otimes\cdots \otimes W^n_u$};
    \node[below] at (0,-1) {$W^1_{d}\otimes W^2_{d} \otimes\cdots \otimes W^n_d$};
    \node[anchor=center, rotate = -90] at (1.2,0) {$W^1_{r}\otimes W^2_{r} \otimes\cdots \otimes W^n_r$};
    \node[anchor=center, rotate = 90] at (-1.2,0) {$W^1_{l}\otimes W^2_{l} \otimes\cdots \otimes W^n_l$};
}
= Z_{\Box}.
\]
Here, $Z_\Box$ is the partition function of the block, which maps
$$ Z_{\Box}: \bigotimes_{i}W^i_l \otimes \bigotimes_{i}W^i_d\leftarrow\bigotimes_{i}W^i_u \otimes \bigotimes_{i}W^i_r.$$
We use the same convention as in TNRKit and TensorKit by reading tensor map orders from up-right to left-down, and the tensor product order from left to right.
In practice, the number of lattice sites $n^2> 3\times 10^4$ when a tensor reaches the fixed-point.
The spaces (legs) $V_u$, $V_r$, $V_l$, $V_d$ of the tensor carry the space of the upper, right, left, and down boundary conditions assigned to the edges of that block respectively\footnote{More rigorously, the fixed-point tensor $T_*$ should be composed with isometries $\Psi_\bullet:V_\bullet\rightarrow \otimes_i W^i_\bullet$ to be comparable with $Z_{\Box}$, see Eq.(9) in \cite{Levin_trg}.}. When several such tensors are contracted, the partition functions for the individual blocks are ``glued'' together, with the boundary conditions summed over.
\[
\diagram[1.2]{
    \draw (-1,0) -- (3,0);
    \draw (0,-1) -- (0,1);
    \draw (2,-1) -- (2,1);
    \filldraw[fill=white, draw = black] (-.2,-.2) rectangle (.2,.2);
    \filldraw[fill=white, draw = black] (2-.2,-.2) rectangle (2.2,.2);
    \node[below] at (1, 0){$V_m$};
} \approx
\diagram[1.2]{
    \foreach \i in {-0.9, -0.8, ..., 0.9}
    {
    \draw[black!20] (-1,\i) -- (3,\i);
    }
    \foreach \j in {-0.9, -0.8, ..., 2.9}
    {
        \draw[black!20] (\j,-1) -- (\j,1);
    }
    \draw[red] (1, 1) -- node[right]{$\otimes_i W^i_m$} (1, -1);
}= \tr_{\otimes_i W^i_m}(Z_{\Box}\otimes Z_{\Box}).
\]

Therefore, there is a correspondence between fixed-point tensors and geometries. As an application of this correspondence, the contraction of two opposite legs of the fixed-point tensor, called a transfer matrix, corresponds to placing the tensor network on a tube (periodic boundary conditions) \cite{Gu_2009}. 
\[
    \diagram[1]{
\draw[string] (0,1) -- (0,0);
\draw[string] (0,0) -- (0,-1);
\draw[string,knot, rounded corners = 5pt] (0,0) -- (-1,0) -- (-1,.5) -- (1,.5) -- (1,0) -- (0,0);
\filldraw[fill=white, draw = black] (-.2,-.2) rectangle node{$T_*$} (.2,.2);
}\approx
Z_{\mathrm{tube}}.
\]
Here $Z_{\mathrm{tube}}$ is the map
\[
Z_{\mathrm{tube}}:\bigotimes_{i}W^i_d\leftarrow\bigotimes_{i}W^i_u.
\]
At the fixed-point, the transfer matrix becomes an imaginary-time evolution operator on a tube of a CFT or TQFT, thus the spectrum of the transfer matrix encodes the universal data of a phase by its energy spectrum and the corresponding eigenstates on a circle \cite{Gu_2009}.

\subsection{Obtaining scaling dimensions}
\begin{figure}[tb]
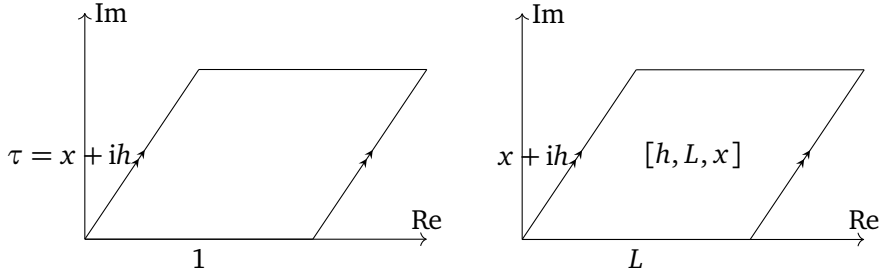

    \centering
    \diagram[1.5]{
\draw[sstring] (-1,-1.5) -- node[left]{$\tau = x + \ii h$} (0, 0);
\draw (0,0) -- (2, 0);
\draw[sstring] (1,-1.5) --  (2, 0);
\draw (-1, -1.5) -- node[below]{$1$} (1,-1.5);
\draw[->] (-1,-1.5) -- (-1,.5) node[right]{$\Im$};
\draw[->] (-1,-1.5) -- (2,-1.5) node[above]{$\Re$};
}\quad 
\diagram[1.5]{
\draw[sstring] (-1,-1.5) -- node[left]{$x + \ii h$} (0, 0);
\draw (0,0) -- (2, 0);
\draw[sstring] (1,-1.5) --  (2, 0);
\draw (-1, -1.5) -- node[below]{$L$} (1,-1.5);
\node at (.5, -.75) {$[h, L, x]$};
\draw[->] (-1,-1.5) -- (-1,.5) node[right]{$\Im$};
\draw[->] (-1,-1.5) -- (2,-1.5) node[above]{$\Re$};
}
    \caption{Geometry of the tube. We use double arrows to indicate the identification (gluing) of opposite edges of a parallelogram. The upper and bottom edges of the parallelogram are the upper and bottom edges of the tube, respectively. The left figure represents a scale-invariant tube. The perimeter of the circle is fixed to unity. The modular parameter $\tau$ lies in the upper half-plane: its imaginary part corresponds to the height $h$ of the tube, while its real part represents the horizontal displacement $x$. The right figure depicts a tube arising from a lattice model, which introduces an additional length scale $L$, corresponding to the perimeter of the circle. The shape of this tube is denoted as $[h, L, x]$.}
    \label{fig: tube geometry}
\end{figure}

If the lattice system is described by a conformal field theory (CFT) in the fixed-point, the evolution operator has a compact form
\[
Z(\tau, \bar\tau) = q^{L_0 - \frac{c}{24}}\bar{q}^{\bar{L}_0 - \frac{c}{24}}:\cH\rightarrow\cH,\quad q = \ee^{2\pi\ii \tau}.
\]
Here $\cH$ is the Hilbert space on a circle, $\tau$ is the modular parameter of the tube (see Fig.~\ref{fig: tube geometry}), $L_0$ ($\bar{L}_0$) is the chiral (anti-chiral) Virasoro energy operator and $c$ is the central charge.
Expanding $\tau = x + \ii h$, the evolution operator can be expressed as
\[
Z(x, \beta) = \ee^{2\pi\ii xP - 2\pi hH}.
\]
Here
\[
H = L_0 + \bar{L}_0 - \frac{c}{12},\quad P = L_0 - \bar{L}_0
\]
are energy and momentum respectively. The operator $\hat{\Delta}:= L_0 + \bar{L}_0$ without central charge is called the scaling dimension operator in CFT, whose eigenvalues $\{\Delta_i\}_i$ are called scaling dimensions. The central charge term $-c/12$ shifts the scaling dimension operator by a constant and is usually referred to as the zero-point energy. The eigenvalues $\{s_i\}_i$ of the translation operator $P$ are called conformal spins. Since $\hat{\Delta}$ and $P$ commute, they can be simultaneously diagonalized, and the pair $(\Delta, s)$ serves as a set of good quantum numbers.
The evolution operator $Z(\tau, \bar{\tau})$ itself is a composition of an imaginary time evolution by $-\ii h$ and a translation by $x$.

The evolution operator in CFT is scale-invariant, only depending on the shape of the tube.
However, different from the purely theoretical aspect, evolution operators obtained from lattice models usually contain a non-universal area contribution \cite{wei2023tensornetworkrenormalizationapplication,Wen_2020, Gu_2009}, proportional to the free energy density:
\[
Z_{\text{lattice}}\left(\frac{x+\ii h}{L}, \frac{x-\ii h}{L}; L\right) = \ee^{f\times hL}q^{L_0 - \frac{c}{24}}\bar{q}^{\bar{L}_0 - \frac{c}{24}},\quad q = \ee^{2\pi\ii \frac{x + \ii h}{L}}.
\]
Here, an additional length scale $L$ is introduced. The geometric meaning of $x$ and $h$ is evident from Fig.~\ref{fig: tube geometry}. The area term $e^{f\times hL}$ makes the partition function not scale invariant exactly, but \emph{up to a scalar} \cite{Gu_2009}. Instead of using $\tau$ alone to label a shape of a tube in CFT, we will use a triple $[h, L, x]$ to denote a \emph{shape} of a tube obtained from the lattice and denote
\[
Z_{[h, L, x]} := Z_{\text{lattice}}\left(\frac{x+\ii h}{L}, \frac{x-\ii h}{L}; L\right).
\]

Note that there exists an ambiguity in choosing the length unit, which is equivalent to defining the unit of the free energy density. We will fix the convention by choosing the area of the fixed-point tensor at the current RG step as $L^2 = 1$.

As an illustrative example, we choose the shape to be $[1, 1, 0]$, which means the tube is now a gluing of a square. For the convention of the shape, see Fig.\ref{fig: tube geometry}. In this case, $q = \ee^{-2\pi}$, and the evolution operator becomes
\begin{equation*}
\vcenter{\hbox{
\diagram[1]{
  \draw[string] (0,1) -- (0,0);
  \draw[string] (0,0) -- (0,-1);
  \draw[string,knot, rounded corners=5pt]
    (0,0) -- (-1,0) -- (-1,.5) -- (1,.5) -- (1,0) -- (0,0);
  \filldraw[fill=white, draw=black] (-.2,-.2) rectangle (.2,.2);
  \node at (0,0) {$T_{*}$};
}}}
\approx Z_{[1,1,0]}
= \ee^{f}\ee^{-2\pi\left(\hat{\Delta}-\frac{c}{12}\right)}.
\end{equation*}
The spectrum $\{\lambda_i\}_i$ of the transfer matrix encodes the spectrum $\{\Delta_i\}_i$ of $\hat{\Delta}$:
\begin{equation}\label{eq: scaling dim}
    \lambda_i \approx \ee^{f}\ee^{-2\pi \left(\Delta_i - \frac{c}{12}\right)}.
\end{equation}
By taking the ratio of eigenvalues, we can get rid of the area term and measure the difference between scaling dimensions:
\[
\Delta_i -\Delta_{\mathrm{min}} \approx -\frac{1}{2\pi}\log \frac{\lambda_i}{\lambda_\mathrm{max}}. 
\]
In unitary theories, the lowest scaling dimension $\Delta_{\mathrm{min}} = 0$. We can thus reconstruct scaling dimensions from eigenvalues of a transfer matrix~\cite{Gu_2009}. In non-unitary theories, such as the Yang-Lee CFT, we obtain effective scaling dimensions $\{\Delta_i -\Delta_{\mathrm{min}}\}_i$ \cite{wei2023tensornetworkrenormalizationapplication,bao2019,bao2025tensorcomplexrenormalizationgeneralized}.

\subsection{Obtaining central charge}
From Eq.~\eqref{eq: scaling dim}, we may infer that the central charge is only a normalization constant, which is mixed with the area term and can never be extracted from a fixed-point tensor alone. In this section, we will introduce a method described in Ref.~\citenum{Wen_2020}, which is similar to the strategy of obtaining the universal topological entanglement entropy from a single ground state \cite{Levin_2006, Kitaev_2006}. By comparing two different geometries, we can get rid of the area term and extract the central charge from the data of a single fixed-point tensor.

We denote the largest eigenvalue of the transfer matrix $Z_{[1, 1, 0]}$ in Eq.~\eqref{eq: scaling dim} as
\[
\lambda_{\mathrm{max}} \approx \ee^{f}\ee^{-2\pi \left(\Delta_{\mathrm{min}} - \frac{c}{12}\right)}.
\]
Now we choose another shape $[m, n, 0]$, where $m$ and $n$ can potentially be any real numbers. In Sec.~\ref {sec:shape}, we will introduce the jigsaw trick to produce different shapes from a single fixed-point tensor. 

The largest eigenvalue of $Z_{[m, n,0]}$ should be of the form
\[
\lambda'_{\mathrm{max}} \approx \ee^{fmn}\ee^{-2\pi\frac{m}{n} \left(\Delta_{\mathrm{min}} - \frac{c}{12}\right)}.
\]
Consider
\[
\frac{\lambda'_\mathrm{max}}{\lambda_\mathrm{max}^{mn}} \approx \ee^{2\pi m\left(n - \frac{1}{n}\right)\left(\Delta_{\mathrm{min}} - \frac{c}{12}\right)},
\]
whenever $n\neq 1$, we obtain
\[
c - 12\Delta_{\mathrm{min}} \approx \frac{12}{2\pi m\left(\frac{1}{n} - n\right)}\log\frac{\lambda_\mathrm{max}'}{\lambda_{\mathrm{max}}^{mn}}.
\]
Again, for unitary theories, $\Delta_{\mathrm{min}} = 0$ and we obtain the central charge \cite{Gu_2009}. For non-unitary theories, we obtain the effective central charge $c - 12 \Delta_{\mathrm{min}}$ \cite{wei2023tensornetworkrenormalizationapplication,bao2019,bao2025tensorcomplexrenormalizationgeneralized}.
Physically, changing $L$ in $[h, L, 0]$ is similar to the Casimir effect, which is one possible way to detect the zero-point energy $-c/12$.

It is impossible to calculate $\Delta_{\mathrm{min}}$ alone, only looking at eigenvalues of the transfer matrix, as there is always the ambiguity of shifting $\hat\Delta + \delta$ and $c/12 + \delta$ simultaneously.

\subsection{Obtaining conformal spins}
Now we choose a shape $[1, 1, x]$, where $x$ can be non-zero. Eigenvalues of $Z_{[1, 1, x]}$ will look like
\[
\lambda_i = \ee^{f}\ee^{-2\pi\left(\Delta_i- \frac{c}{12}\right)}\ee^{2\pi \ii xs_i}.
\]
Taking the quotient, we have
\[
\frac{\lambda_i}{\lambda_{\mathrm{max}}} = \ee^{-2\pi(\Delta_i - \Delta_{\mathrm{min}})}\ee^{2\pi\ii xs_i}.
\]
Here, we have assumed the conformal spin of the ground state is zero, which is true for most unitary theories and non-unitary theories. Scaling dimensions $\{\Delta_i\}_i$ and conformal spins $\{s_i\}_i$ are encoded in absolute values and phase factors of $\{\lambda_i/\lambda_{\mathrm{max}}\}_i$.

For a purely two-dimensional bosonic system, a horizontal translation by $x=1$ acts as the identity operator. This implies the constraint
\[
\ee^{2\pi \ii P} = 1.
\]
Consequently, all conformal spins must be integer-valued:
\[
s_i \in \mathbb{Z}.
\]
This condition is equivalently expressed as modular invariance under the $T$-transformation when considering the torus partition function, rather than the evolution operator:
\[
Z(\tau+1, \bar{\tau}+1) = Z(\tau, \bar{\tau}).
\]
The constraint on conformal spins is discrete and robust under perturbations, which makes conformal spins typically more accurate than scaling dimensions in numerical computations.

When solving $s_i$ from $\ee^{2\pi \ii x s_i}$, logarithm of a complex number is non-unique:
\[
xs_i \in \frac{1}{2\pi}\arg \frac{\lambda_i}{\lambda_{\mathrm{max}}} + \Z.
\]
When $x = p/q$ with $p,q\in\Z$ being coprime, $s_i$ can only be fixed up to $q$ \cite{Hauru_2016}. The final data that can be extracted from a fixed-point tensor is
\[
\bigg\{\big(\Delta_i - \Delta_{\min},\ s_i(\mathrm{mod}\,q)\big)\bigg\}_i,\quad c - 12\Delta_{\min}.
\]

Note that when computing $(1+1)$D quantum models, for example, in Refs.~\citenum{bao2019} and \citenum{wei2023tensornetworkrenormalizationapplication}, it is possible to continuously deform the modular parameter $\tau$ by continuously changing the initial time scale $\delta\beta$ in the Trotter expansion. Therefore, it is possible to perform the derivative to $x$ in $\ee^{2\pi \ii xs_i}$, producing an exact conformal spin without the $\mathrm{mod}\, q$-ambiguity. Such a continuous deformation so far does not exist in classical models.

\subsection{Overcoming finite bond dimension effects and resolving higher descendants}
If the tensor network reaches the exact fixed-point, because of scale invariance, taking different shapes of transfer matrices will be equivalent for resolving scaling dimensions. However, finite bond dimension makes the fixed-point only an approximation \cite{FiniteD_2014, FiniteD_2023}. In CFT, an infinite-dimensional Hilbert space on a circle is needed to host all descendant states. For example, the Hilbert space of the Ising CFT on a circle is
\[
\cH = \cV_0\otimes \overline{\cV_0} \oplus \cV_{\frac{1}{2}}\otimes \overline{\cV_{\frac{1}{2}}} \oplus \cV_{\frac{1}{16}}\otimes \overline{\cV_{\frac{1}{16}}}.
\]
Each $\cV_{0}$, $\cV_{\frac{1}{2}}$ and $\cV_{\frac{1}{16}}$ are chiral conformal families containing infinitely-many descendants. In contrast, the finite bond dimension of a fixed-point tensor is unable to host all states in $\cH$. Therefore, different shapes of transfer matrices will have different performances in practice.

States with different energy levels contribute unequally at a finite bond dimension. From the expression of the torus partition function at $\tau = \ii$, see Fig.~\ref{fig:tracenorm},
\[
\sum_{i} \ee^{-2\pi \left(\Delta_i - c/12\right)},
\]
we see that the contributions decay exponentially with $\Delta_i$. Therefore, when approximating local tensors, descendant states are more strongly suppressed and thus play a less significant role in a finite-dimensional Hilbert space. Consequently, even though primary states can be computed accurately in Ref.~\citenum{Gu_2009}, states with higher scaling dimensions remain poorly resolved.

As has been discovered in Ref.~\citenum{Looptnr_2017}, taking a wider transfer matrix, that is, increasing $L$ in $[h, L, x]$, can help resolve higher descendants.
Intuitively, taking larger patches of a tube will produce larger effective bond dimensions. Suppose the bond dimension of a fixed-point tensor with shape $[1, 1, 0]$ is $\chi$, then choosing a tube with shape $[2, 2, 0]$ will effectively increase the dimension of the Hilbert space to $\chi^2$:
\[
\diagram[1]{
\draw (0,1) -- (0,0);
\draw (0,0) -- (0,-1);
\draw[knot, rounded corners = 5pt] (0,0) -- (-1,0) -- (-1,.5) -- (1,.5) -- (1,0) -- (0,0);
\filldraw[fill=white, draw = black] (-.2,-.2) rectangle  (.2,.2);
} \leadsto
\diagram[1]{
    \draw (0,-3) -- (0,1);
    \draw (2,-3) -- (2,1);
    \draw[knot, rounded corners = 5pt] (-1,-2) -- (3,-2) -- (3, -2+.5) -- (-1, -2+.5) -- cycle;
    \draw[knot, rounded corners = 5pt] (-1,0) -- (3,0) -- (3,.5) -- (-1,.5) -- cycle;
    \filldraw[fill=white, draw = black] (-.2,-.2) rectangle (.2,.2);
    \filldraw[fill=white, draw = black] (-.2,-2-.2) rectangle (.2,-2+.2);
    \filldraw[fill=white, draw = black] (2-.2,-.2) rectangle (2.2,.2);
    \filldraw[fill=white, draw = black] (2-.2,-2-.2) rectangle (2.2,-2+.2);
}.
\]
Suppose $\chi = 16$, then $\chi^2 = 256$, which greatly improves the representability of descendants.

The resolution of higher descendant states can also be understood from the perspective of the RG flow. If the discarded operators correspond to irrelevant perturbations of a CFT, then coarse-graining over more tensors drives the system back toward the fixed point. Increasing $L$ can thus be interpreted as an RG flow that suppresses irrelevant perturbations.

We emphasize that increasing $h$ does not help resolve descendants. One reason is that increasing $h$ does not increase the dimension of the Hilbert space. Even worse, because the spectrum of the transfer matrix is determined by the ratio $h/L$: 
\[\frac{\lambda_i}{\lambda_{\mathrm{max}}} = \ee^{-2\pi\frac{h}{L}\Delta_i},\]
a larger modular parameter makes the eigenvalue of the transfer matrix decay faster.
The lesson is therefore to make $L$ larger while keeping $h$ small. For example at $\Delta=4$, $\ee^{-2\pi\Delta} \sim 10^{-11}$ while $\ee^{-2{\frac{1}{2}\pi\Delta}} \sim 10^{-5}$. Tiny numbers are notoriously vulnerable to numerical noise, so this is another reason why a wider transfer matrix is beneficial for stability.

\subsection{Generating other shapes: playing with a jigsaw puzzle}\label{sec:shape}
By the arguments of the last section, one might wish to take L as large as possible; however, doing so is computationally costly. The problem therefore becomes how to construct a transfer matrix with a large L while keeping the computational cost manageable. Since both Levin–Nave TRG and LoopTNR have a cost of \(\mathcal{O}(\chi^6)\), it is desirable to solve for the eigenvalues of the transfer matrix with the same cost.

First, we can apply ideas in exact diagonalization by avoiding forming a dense transfer matrix and only defining a linear action function of that transfer matrix on an arbitrary state. Having that action function, we use Arnoldi iteration to approximately solve the leading eigenvalues of a transfer matrix. In TNRKit.jl, we use the Arnoldi iteration function provided in KrylovKit.jl \cite{Haegeman_2025_KrylovKit} to solve for the leading eigenvalues, which can be applied to both systems with and without symmetries. Since a transfer matrix can be defined as a contraction of several smaller tensors, the action function of the transfer matrix can also be decomposed into several cheaper tensor contractions. Therefore, using Arnoldi iteration can greatly enlarge the possible $L$ we can achieve \cite{Looptnr_2017, bao2019}.

The transfer matrix with shape $[2, 4, 0]$, as introduced in Ref.~\citenum{Looptnr_2017}, is:
\[
\diagram[.7]{
\foreach \i in {0, 2, 4, 6}
{
\draw (\i,-3) -- (\i,1);
}
\draw[knot, rounded corners = 2.5pt] (-1, 0) -- (7,0) -- (7,.5) -- (-1,.5) -- cycle;
\draw[knot, rounded corners = 2.5pt] (-1, -2) -- (7,-2) -- (7,-2+.5) -- (-1,.5-2) -- cycle;
\foreach \i in {0, 4}
{
\filldraw[fill = white, draw = black] (\i-.2, -.2) rectangle (\i+.2,.2);
\filldraw[fill = black, draw = black] (\i-.2, -2-.2) rectangle (\i+.2,-2+.2);
}
\foreach \i in {2, 6}
{
\filldraw[fill = black, draw = black] (\i-.2, -.2) rectangle (\i+.2,.2);
\filldraw[fill = white, draw = black] (\i-.2, -2-.2) rectangle (\i+.2,-2+.2);
}
} = 
\diagram[.7]{
\foreach \i in {0, 4}
{
\fill[black!20] (\i-1, -1) rectangle (\i+1,1);
\fill[black!40] (\i-1, -2-1) rectangle (\i+1,-2+1);
}
\foreach \i in {2, 6}
{
\fill[black!20] (\i-1, -2-1) rectangle (\i+1,-2+1);
\fill[black!40] (\i-1, -1) rectangle (\i+1,+1);
}
\draw[sstring] (-1,-3) -- (-1,1);
\draw[sstring] (7,-3) -- (7,1);
}
\]
Contracting and storing such a dense transfer matrix will be very expensive. Instead, in Ref.~\citenum{Looptnr_2017}, using Arnoldi iteration, the action function is defined as
\[
\diagram[.4]{
\foreach \i in {0, 2, 4, 6}
{
\draw (\i, -3) -- (\i,-2);
}
\foreach \i in {0}
{
\draw (\i, -3) -- (\i,-1);
\draw (\i - 1, -2) -- (\i + 1,-2);
\filldraw[fill = white, draw = black] (\i-.2, -2-.2) rectangle (\i+.2,-2+.2);
}
\filldraw[fill = white, draw = black] (-.2, -3.2) rectangle node[below]{$\cO(\chi^7)$} (6.2, -2.8);
}\rightarrow
\diagram[.4]{
\foreach \i in {0, 2, 4, 6}
{
\draw (\i, -3) -- (\i,-2);
}
\foreach \i in {0, 2}
{
\draw (\i, -3) -- (\i,-1);
\draw (\i - 1, -2) -- (\i + 1,-2);
\filldraw[fill = white, draw = black] (\i-.2, -2-.2) rectangle (\i+.2,-2+.2);
}
\filldraw[fill = white, draw = black] (-.2, -3.2) rectangle node[below]{$\cO(\chi^8)$} (6.2, -2.8);
}
\rightarrow
\diagram[.4]{
\foreach \i in {0, 2, 4, 6}
{
\draw (\i, -3) -- (\i,-2);
}
\foreach \i in {0, 2, 4}
{
\draw (\i, -3) -- (\i,-1);
\draw (\i - 1, -2) -- (\i + 1,-2);
\filldraw[fill = white, draw = black] (\i-.2, -2-.2) rectangle (\i+.2,-2+.2);
}
\filldraw[fill = white, draw = black] (-.2, -3.2) rectangle node[below]{$\cO(\chi^8)$} (6.2, -2.8);
}
\rightarrow
\diagram[.4]{
\foreach \i in {0, 2, 4, 6}
{
\draw (\i, -3) -- (\i,-1);
\filldraw[fill = white, draw = black] (\i-.2, -2-.2) rectangle (\i+.2,-2+.2);
}
\draw[knot, rounded corners = 1.5pt] (-1, -2) -- (7,-2) -- (7,-2+.5) -- (-1,.5-2) -- cycle;
\foreach \i in {0, 2, 4, 6}
{
\filldraw[fill = white, draw = black] (\i-.2, -2-.2) rectangle (\i+.2,-2+.2);
}
\filldraw[fill = white, draw = black] (-.2, -3.2) rectangle node[below]{$\cO(\chi^7)$} (6.2, -2.8);
},
\]
bringing the contraction cost in the Arnoldi iteration to $\cO(4\chi^8)$. Even though the computational cost is already much cheaper than the dense approach, it is still more expensive than the $\cO(\chi^6)$ of LoopTNR. Using this transfer matrix, one can at most contract tensors with bond dimension $\chi = 16$ on a desktop computer within 10 minutes.

Note that in Ref.~\citenum{Looptnr_2017}, the transfer matrix is chosen to be two layers, i.e. $h = 2$, only because there are two types of tensors in LoopTNR. We can slightly improve this transfer matrix by only considering a single layer, i.e. $h = 1$, and introducing a translation by $x = 1$ \cite{Hauru_2016}. Therefore, we reduce the computational cost to $\cO(2\chi^8)$ and simultaneously gain a resolution of conformal spins modulo $4$. The shape of this geometry is $[1, 4, 1]$:
\begin{equation}\label{eq:transfer origin}
    \diagram[.7]{
\foreach \i in {0, 2, 4}
{
\draw[rounded corners = 2.5pt] (\i + 2,-1.5) -- (\i + 2,-1.2) -- (\i, -1) --  (\i,1);
}
\draw[knot, rounded corners = 2.5pt] (-1, 0) -- (7,0) -- (7,.5) -- (-1,.5) -- cycle;
\draw[knot, rounded corners = 2.5pt] (0,-1.5) -- (0,-1.2) -- (-1, -1) --  (-1,-.5) -- (7, -.5) -- (7, -1) -- (6,-1) -- (6,1);
\foreach \i in {0, 4}
{
\filldraw[fill = white, draw = black] (\i-.2, -.2) rectangle (\i+.2,.2);
}
\foreach \i in {2, 6}
{
\filldraw[fill = black, draw = black] (\i-.2, -.2) rectangle (\i+.2,.2);
}
} = 
\diagram[.7]{
\filldraw[fill = black!40, dashed] (-3, -1) -- (-1, -1) -- (-1, 1) -- cycle;
\foreach \i in {0, 4}
{
\fill[black!20] (\i-1, -1) rectangle (\i+1,1);
}
\foreach \i in {2, 6}
{
\fill[black!40] (\i-1, -1) rectangle (\i+1,+1);
}
\draw[dashed] (5, -1) -- (7, -1) -- (7, 1) -- cycle;
\draw[sstring] (-1,-1) -- (-1,1);
\draw[sstring] (7,-1) -- (7,1);
\draw[ssstring, white] (-1.5,-.5) -- (6.5,-.5);
}.
\end{equation}

\subsubsection{The horizontal jigsaw trick}
We should take a closer look and get a better understanding of Eq.~\eqref{eq:transfer origin}. At first sight, if we need a transfer matrix with shape $[1, 4, 1]$, we need the following steps:
\begin{enumerate}
    \item Decompose each fixed-point rank-4 tensor into two rank-3 tensors via exact SVD (c.f. Eq.~\eqref{eq:TRG-svd}). Geometrically, this decomposition is equivalent to decomposing a partition function on a square into a gluing of that on two triangles. Different from the SVD Eq.~\eqref{eq:TRG-svd} in TRG, we emphasize that here we do not truncate the singular values.
\begin{align}\label{eq:SVD geometry}
        \diagram[.8]{
        \draw (0,1) -- (0,-1);
\draw (1, 0) -- (-1, 0);
\filldraw[fill = white, draw = black] (-.2,-.2) rectangle (.2, .2);
}
&\overset{\text{SVD}}{=}
\diagram[.8]{
\draw (-1, 0) -- (0,0);
\draw (0, 0) -- (0,1);
\draw[ultra thick] (0,0) -- node[above right]{$\chi^2$} (1, -1);
\draw (1, -1) -- (1, -2);
\draw (1, -1) -- (2, -1);
\filldraw[fill = white, draw = black] (0,0) circle (.2);
\filldraw[fill = white, draw = black] (1,-1) circle (.2);
} \Leftrightarrow 
\diagram[.8]{
\fill[black!20] (-1, -1) rectangle (1, 1);
}
=
\diagram[.8]{
\fill[black!20] (-1, -1) -- (1, 1) -- (-1, 1) -- cycle;
\fill[black!20] (-1+.2, -1-.2) -- (1+.2, 1-.2) -- (1+.2, -1-.2) -- cycle;
\draw[sstring] (-1, -1) -- (1, 1);
\draw[sstring] (-1+.2, -1-.2) -- (1+.2, 1-.2);
},\nonumber\\
\diagram[.8]{
\draw (0,1) -- (0,-1);
\draw (1, 0) -- (-1, 0);
\filldraw[fill = black, draw = black] (-.2,-.2) rectangle (.2, .2);
}
&\overset{\text{SVD}}{=}
\diagram[.8]{
\draw (-1, 0) -- (0,0);
\draw (0, 0) -- (0,1);
\draw[ultra thick] (0,0) -- node[above right]{$\chi^2$} (1, -1);
\draw (1, -1) -- (1, -2);
\draw (1, -1) -- (2, -1);
\filldraw[fill = black, draw = black] (0,0) circle (.2);
\filldraw[fill = black, draw = black] (1,-1) circle (.2);
} \Leftrightarrow 
\diagram[.8]{
\fill[black!40] (-1, -1) rectangle (1, 1);
}
=
\diagram[.8]{
\fill[black!40] (-1, -1) -- (1, 1) -- (-1, 1) -- cycle;
\fill[black!40] (-1+.2, -1-.2) -- (1+.2, 1-.2) -- (1+.2, -1-.2) -- cycle;
\draw[sstring] (-1, -1) -- (1, 1);
\draw[sstring] (-1+.2, -1-.2) -- (1+.2, 1-.2);
}.
\end{align}
\item From the new rank-3 tensors, construct two new rank-4 tensors by contraction. Geometrically, this step is equivalent to forming two parallelograms with shape $[1, 1, 1]$:
\begin{align*}
    \diagram[.8]{
\draw (0,1) -- (0,-1);
\draw[ultra thick] (1, 0) -- (-1, 0);
\filldraw[fill = black, draw = black] (.2*0.707, -.2*0.707) arc (-45:45+90:.2) -- cycle;
\filldraw[fill = white, draw = black] (.2*0.707, -.2*0.707) arc (-45:-45-180:.2) -- cycle;
}
&:=
\diagram[.8]{
\draw[ultra thick] (0,0) -- (-1, 1);
\draw (0,0) -- (0,-1);
\draw (0,0) -- (1, 0);
\draw (1,0) -- (1, 1);
\draw[ultra thick] (1,0) -- (2, -1);
\filldraw[fill = white, draw = black] (0,0) circle (.2);
\filldraw[fill = black, draw = black] (1,0) circle (.2);
}\Leftrightarrow
\diagram[1]{
\fill[black!20] (-1, 0) -- (0,0) -- (0,1) -- cycle;
\fill[black!40] (1, 1) -- (0,1) -- (0,0) -- cycle;
},\\
\diagram[.8]{
\draw (0,1) -- (0,-1);
\draw[ultra thick] (1, 0) -- (-1, 0);
\filldraw[fill = white, draw = black] (.2*0.707, -.2*0.707) arc (-45:45+90:.2) -- cycle;
\filldraw[fill = black, draw = black] (.2*0.707, -.2*0.707) arc (-45:-45-180:.2) -- cycle;
}
&:=
\diagram[.8]{
\draw[ultra thick] (0,0) -- (-1, 1);
\draw (0,0) -- (0,-1);
\draw (0,0) -- (1, 0);
\draw (1,0) -- (1, 1);
\draw[ultra thick] (1,0) -- (2, -1);
\filldraw[fill = black, draw = black] (0,0) circle (.2);
\filldraw[fill = white, draw = black] (1,0) circle (.2);
}\Leftrightarrow
\diagram[1]{
\fill[black!40] (-1, 0) -- (0,0) -- (0,1) -- cycle;
\fill[black!20] (1, 1) -- (0,1) -- (0,0) -- cycle;
};
\end{align*}
\item Define a new transfer matrix from four new rank-4 tensors, geometrically correspond to a $[1, 4, 1]$ tube:
\begin{equation}\label{eq:transfer new}
    \diagram[.8]{
\foreach \i in {0, 2, 4, 6}
{
\draw (\i,1) -- (\i, -1);
}
\draw[knot, rounded corners = 2.5pt, ultra thick] (-1, 0) -- (7,0) -- (7,.5) -- (-1,.5) -- cycle;
\foreach \i in {0, 4}
{
\filldraw[fill = white, draw = black] (\i+.2*0.707, -.2*0.707) arc (-45:45+90:.2) -- cycle;
\filldraw[fill = black, draw = black] (\i+.2*0.707, -.2*0.707) arc (-45:-45-180:.2) -- cycle;
}
\foreach \i in {2, 6}
{
\filldraw[fill = black, draw = black] (\i+.2*0.707, -.2*0.707) arc (-45:45+90:.2) -- cycle;
\filldraw[fill = white, draw = black] (\i+.2*0.707, -.2*0.707) arc (-45:-45-180:.2) -- cycle;
}
}.
\end{equation}
\end{enumerate}
If we directly use the transfer matrix \eqref{eq:transfer new}, the contraction cost in the Arnoldi iteration will rise to $\cO(2\chi^{11})$ because the dense SVD doubles virtual bond dimensions of the MPO. Luckily, by shuffling the contraction order and contracting back two rank-3 tensors into the original rank-4 tensor, we find the transfer matrix \eqref{eq:transfer new} is exactly equivalent to the one constructed in \eqref{eq:transfer origin}. The shuffling rule of a rank-3 tensor is indicated by the white arrow in \eqref{eq:transfer origin}. This is the reason why $\cO(2\chi^8)$ will be enough for the $[1, 4, 1]$ transfer matrix.

In general, by shuffling tensor contraction orders, one may avoid forming tensors with large bond dimensions in dense SVD. Geometrically, the shuffling of tensor contraction orders corresponding to the shuffling of gluing orders of partition functions on triangles. We call this trick the \emph{jigsaw trick}. This trick already exists in previous literatures for example in Refs.~\citenum{bao2019} and~\citenum{Hauru_2016}, but we will develop it more systematically here.

\subsubsection{The vertical jigsaw trick}
In the previous section, we introduced the jigsaw trick that shuffles triangles \emph{horizontally}, which produces an \emph{equivalent} transfer matrix.
Now we consider the $[\sqrt{2}, 2\sqrt{2}, 0]$ transfer matrix and introduce the idea of the \emph{vertical} jigsaw trick along the way. Different from the horizontal jigsaw trick, the vertical jigsaw trick may change the transfer matrix while only keeping its \emph{spectrum} (and degeneracies) invariant.

A naive $[\sqrt{2}, 2\sqrt{2},0]$ transfer matrix is
\[
\diagram[.8]{
\draw[ultra thick] (0,0) -- (0,1);
\draw[ultra thick] (2,0) -- (2,1);
\draw[ultra thick] (0,-2) -- (0,-3);
\draw[ultra thick] (2,-2) -- (2,-3);
\draw (0,0) -- (-1, -1) -- (0, -2) -- (2, 0) -- (3, -1) -- (2, -2) -- cycle;
\draw[knot, rounded corners = 2.5pt, ultra thick] (-1, -1) -- (-2, -1) -- (-2, -.5) -- (4, -.5) -- (4,-1) -- (3, -1);
\filldraw[fill = white] (0,0) circle (.2);
\filldraw[fill = white] (0,-2) circle (.2);
\filldraw[fill = white] (2,0) circle (.2);
\filldraw[fill = white] (2,-2) circle (.2);
\filldraw[fill = black] (-1,-1) circle (.2);
\filldraw[fill = black] (3,-1) circle (.2);
\filldraw[fill = black] (1, -1+.3) -- (1-.3, -1) -- (1, -1 - .3) -- (1.3, -1) -- cycle;
} \Leftrightarrow
\diagram[1]{
\fill[black!20] (0,0) -- (1, -1) -- (2, 0) -- cycle;
\fill[black!40] (0,0) -- (1, -1) -- (0, -2) -- cycle;
\fill[black!40] (2,0) -- (1, -1) -- (2, -2) -- (3, -1) -- cycle;
\fill[black!20] (0,-2) -- (1, -1) -- (2, -2) -- cycle;
\fill[black!20] (2,0) -- (3,-1) -- (4, 0) -- cycle;
\fill[black!40] (4,0) -- (3, -1) -- (4, -2) -- cycle;
\fill[black!20] (2, -2) -- (3, -1) -- (4, -2) -- cycle;
\draw[sstring] (0,-2) -- (0,0);
\draw[sstring] (4, -2) -- (4, 0);
}.
\]
Here all conventions are the same as Eq.~\eqref{eq:SVD geometry}. Applying the horizontal jigsaw trick, the transfer matrix can be simplified into
\[
\diagram[.8]{
\draw (0,0) -- (2, -2) -- (3,-1) -- (2, 0) -- (0, -2);
\draw[knot, rounded corners = 3pt] (0,0) -- (-.5,-.5) -- (3.5, -.5) -- (3, -1);
\draw[knot, rounded corners = 3pt] (0,-2) -- (-.5,-1.5) -- (3.5, -1.5) -- (3, -1);
\draw[ultra thick] (0,0) -- (0,1);
\draw[ultra thick] (2,0) -- (2,1);
\draw[ultra thick] (0,-2) -- (0,-3);
\draw[ultra thick] (2,-2) -- (2,-3);
\filldraw[fill = white] (0,0) circle (.2);
\filldraw[fill = white] (0,-2) circle (.2);
\filldraw[fill = white] (2,0) circle (.2);
\filldraw[fill = white] (2,-2) circle (.2);
\filldraw[fill = black] (1, -1+.3) -- (1-.3, -1) -- (1, -1 - .3) -- (1.3, -1) -- cycle;
\filldraw[fill = black] (3, -1+.3) -- (3-.3, -1) -- (3, -1 - .3) -- (3.3, -1) -- cycle;
}
\Leftrightarrow
\diagram[1]{
\fill[black!20] (0,0) -- (1, -1) -- (2, 0) -- cycle;
\fill[black!40] (2,0) -- (1, -1) -- (2, -2) -- (3, -1) -- cycle;
\fill[black!20] (0,-2) -- (1, -1) -- (2, -2) -- cycle;
\fill[black!20] (2,0) -- (3,-1) -- (4, 0) -- cycle;
\fill[black!40] (4,0) -- (3, -1) -- (4, -2) -- (5,-1) -- cycle;
\fill[black!20] (2, -2) -- (3, -1) -- (4, -2) -- cycle;
\draw[sstring, red] (1,-1) -- (0,0);
\draw[sstring, blue] (0, -2) -- (1, -1);
\draw[sstring, red] (5,-1) -- (4,0);
\draw[sstring, blue] (4, -2) -- (5, -1);
}.
\]
The cost of the action function is $\cO(6\chi^6)$ now. Next, we use a basic linear algebra fact, the proof of which is in Appendix~\ref{sect:AB BA}. If $A$ and $B$ are two matrices with the same dimension, then the characteristic polynomial (encoding the data of the spectrum and degeneracies) of $AB$, denoted as $p_{AB}(\lambda)$, is the same as $p_{BA}(\lambda)$.

We treat
\[
A = \diagram[.8]{
\draw[ultra thick] (0,0) -- (0,1);
\draw[ultra thick] (2,0) -- (2,1);
\draw (-.5,1.5) -- (0,1) -- (.5,1.5);
\draw (2-.5,1.5) -- (2,1) -- (2.5,1.5);
\filldraw[fill = white] (0,1) circle (.2);
\filldraw[fill = white] (2,1) circle (.2);
},
\]
and $B$ be the rest of the network. Then $p_{AB}(\lambda) = p_{BA}(\lambda)$ implies
\[
p\left(
\diagram[.8]{
\draw (0,0) -- (2, -2) -- (3,-1) -- (2, 0) -- (0, -2);
\draw[knot, rounded corners = 3pt] (0,0) -- (-.5,-.5) -- (3.5, -.5) -- (3, -1);
\draw[knot, rounded corners = 3pt] (0,-2) -- (-.5,-1.5) -- (3.5, -1.5) -- (3, -1);
\draw[ultra thick] (0,0) -- (0,1);
\draw[ultra thick] (2,0) -- (2,1);
\draw[ultra thick] (0,-2) -- (0,-3);
\draw[ultra thick] (2,-2) -- (2,-3);
\filldraw[fill = white] (0,0) circle (.2);
\filldraw[fill = white] (0,-2) circle (.2);
\filldraw[fill = white] (2,0) circle (.2);
\filldraw[fill = white] (2,-2) circle (.2);
\filldraw[fill = black] (1, -1+.3) -- (1-.3, -1) -- (1, -1 - .3) -- (1.3, -1) -- cycle;
\filldraw[fill = black] (3, -1+.3) -- (3-.3, -1) -- (3, -1 - .3) -- (3.3, -1) -- cycle;
}
\right)(\lambda) =
p\left(
\diagram[1]{
\draw[rounded corners = 5pt] (1, 0) -- (1, .5) -- (3, 2.5);
\draw[rounded corners = 5pt] (2, 0) -- (2, .5) -- (0, 2.5);
\draw[knot, rounded corners = 5pt] (3, 0) -- (3, .5) -- (4, 1.5) -- (0,1.5) -- (1, 2.5);
\draw[knot, rounded corners = 5pt] (0,0) -- (0,.5) -- (4,.5) -- (2, 2.5);
\filldraw[fill = black] (1.5, 1+.2) -- (1.5-.2, 1) -- (1.5, 1 - .2) -- (1.5+.2, 1) -- cycle;
\filldraw[fill = black] (3.5, 1+.2) -- (3.5-.2, 1) -- (3.5, 1 - .2) -- (3.5+.2, 1) -- cycle;
\filldraw[fill = white] (.5, 2+.2) -- (.5-.2, 2) -- (.5, 2 - .2) -- (.5+.2, 2) -- cycle;
\filldraw[fill = white] (2.5, 2+.2) -- (2.5-.2, 2) -- (2.5, 2 - .2) -- (2.5+.2, 2) -- cycle;
}
\right)(\lambda).
\]
By reshaping the contraction order vertically, the exact SVD step is avoided. Geometrically we move two triangles upwards to form a zig-zag-shaped boundary.
The new transfer matrix
\begin{equation}\label{eq:two gates}
    \diagram[1.3]{
\draw[rounded corners = 5pt] (1, 0) -- (1, .5) -- (3, 2.5);
\draw[rounded corners = 5pt] (2, 0) -- (2, .5) -- (0, 2.5);
\draw[knot, rounded corners = 5pt] (3, 0) -- (3, .5) -- (4, 1.5) -- (0,1.5) -- (1, 2.5);
\draw[knot, rounded corners = 5pt] (0,0) -- (0,.5) -- (4,.5) -- (2, 2.5);
\filldraw[fill = black] (1.5, 1+.2) -- (1.5-.2, 1) -- (1.5, 1 - .2) -- (1.5+.2, 1) -- cycle;
\filldraw[fill = black] (3.5, 1+.2) -- (3.5-.2, 1) -- (3.5, 1 - .2) -- (3.5+.2, 1) -- cycle;
\filldraw[fill = white] (.5, 2+.2) -- (.5-.2, 2) -- (.5, 2 - .2) -- (.5+.2, 2) -- cycle;
\filldraw[fill = white] (2.5, 2+.2) -- (2.5-.2, 2) -- (2.5, 2 - .2) -- (2.5+.2, 2) -- cycle;
}\Leftrightarrow
\diagram[1]{
\fill[black!20] (0,0) -- (1, -1) -- (2, 0) -- (1, 1) -- cycle;
\fill[black!40] (2,0) -- (1, -1) -- (2, -2) -- (3, -1) -- cycle;
\fill[black!20] (0,-2) -- (1, -1) -- (2, -2) -- cycle;
\fill[black!20] (2,0) -- (3,-1) -- (4, 0) -- (3, 1) -- cycle;
\fill[black!40] (4,0) -- (3, -1) -- (4, -2) -- (5,-1) -- cycle;
\fill[black!20] (2, -2) -- (3, -1) -- (4, -2) -- cycle;
\draw[sstring, red] (1,-1) -- (0,0);
\draw[sstring, blue] (0, -2) -- (1, -1);
\draw[sstring, red] (5,-1) -- (4,0);
\draw[sstring, blue] (4, -2) -- (5, -1);
\draw[dashed] (0,-2) -- (1, -1) -- (2, -2) -- cycle;
\draw[dashed] (2,-2) -- (3, -1) -- (4, -2) -- cycle;
\draw[dashed] (0,0) -- (1, 1) -- (2, 0) -- cycle;
\draw[dashed] (2,0) -- (3, 1) -- (4, 0) -- cycle;
\draw[ssstring, white] (1, -1.5) -- (1, .5);
\draw[ssstring, white] (3, -1.5) -- (3, .5);
}
\end{equation}
originally appears in Ref.~\citenum{bao2019} and has the cost $\cO(4\chi^6)$ in the action function, which can produce the same spectrum (and degeneracies) as the naive one. The transfer matrix \eqref{eq:two gates} is useful both in resolving higher descendants and reducing contraction cost.

\subsection{Resolving higher spins}
The tricks introduced so far allow for the following transformations to shapes:
\begin{itemize}
    \item Gluing: we may contract several rank-4 tensors and rank-3 tensors in horizontal and vertical directions, geometrically corresponding to gluing parallelograms and triangles;
    \item Diagonal decomposition: a rank-4 tensor may be decomposed into two rank-3 tensors through SVD, geometrically corresponding to cutting a parallelogram along its diagonal.
\end{itemize}
These two transformations allow two edges of a parallelogram sliding on the $\Z^2$ lattice:
\[
\diagram[1]{
\fill[black!20] (1,1) -- (2, 3) -- (5,4) -- (4,2) -- cycle;
\foreach \i in {0, 1, ..., 5}
{
    \foreach \j in {0, 1, ..., 5}
    {
    \fill[black] (\i,\j) circle (.05);
    }
}
}.
\]
Modular parameters therefore are all of the form
\[
\tau = \frac{a + b\ii}{c + d \ii},\quad a, b, c, d \in \Z.
\]
The real part of $\tau$, which determines the conformal spin, is always a rational number $p/q$, where $p$ and $q$ are coprime integers. To resolve higher conformal spins, one therefore needs to choose a more suitable unit cell such that $q$ is sufficiently large. However, a large $q$ typically requires a very wide transfer matrix, leading to a high computational cost. In  Ref.~\citenum{Hauru_2016}, a technique was introduced to compress the transfer matrix along the horizontal direction; this approach has also been used, for example, in Ref.~\citenum{bao2025tensorcomplexrenormalizationgeneralized}. The basic idea is similar to TNR by performing a low-rank approximation to a wide transfer matrix.

In this section, we introduce the $\left[\frac{4}{\sqrt{10}}, 2\sqrt{10}, \frac{2}{\sqrt{10}}\right]$ transfer matrix, which can resolve conformal spins modulo $10$. 

Starting from two fixed-point tensors, we can form a new tensor defined on a parallelogram $[1, 2, -1]$ by the following steps:
\begin{enumerate}
    \item Perform two TRG-like SVDs:
    \begin{align*}
        \diagram[.8]{
        \draw (0,1) -- (0,-1);
\draw (1, 0) -- (-1, 0);
\filldraw[fill = white, draw = black] (-.2,-.2) rectangle (.2, .2);
}
&\overset{\text{SVD}}{\approx}
\diagram[.8]{
\draw (-1, 0) -- (0,0);
\draw (0, 0) -- (0,1);
\draw (0,0) -- node[above right]{$\chi$} (1, -1);
\draw (1, -1) -- (1, -2);
\draw (1, -1) -- (2, -1);
\filldraw[fill = white, draw = black] (0,0) circle (.2);
\filldraw[fill = white, draw = black] (1,-1) circle (.2);
} \Leftrightarrow 
\diagram[.8]{
\fill[black!20] (-1, -1) rectangle (1, 1);
}
\approx
\diagram[.8]{
\fill[black!20] (-1, -1) -- (1, 1) -- (-1, 1) -- cycle;
\fill[black!20] (-1+.2, -1-.2) -- (1+.2, 1-.2) -- (1+.2, -1-.2) -- cycle;
\draw[sstring] (-1, -1) -- (1, 1);
\draw[sstring] (-1+.2, -1-.2) -- (1+.2, 1-.2);
},\\
\diagram[.8]{
 \draw (0,1) -- (0,-1);
\draw (1, 0) -- (-1, 0);
\filldraw[fill = black, draw = black] (-.2,-.2) rectangle (.2, .2);
}
&\overset{\text{SVD}}{\approx}
\diagram[.8]{
\draw (1, 0) -- (0,0) -- (0,1);
\draw (0,0) -- node[above left]{$\chi$} (-1, -1);
\draw (-1, -2) -- (-1, -1) -- (-2, -1);
\filldraw[fill = black, draw = black] (0,0) circle (.2);
\filldraw[fill = black, draw = black] (-1,-1) circle (.2);
} \Leftrightarrow 
\diagram[.8]{
\fill[black!40] (-1, -1) rectangle (1, 1);
}
\approx
\diagram[.8]{
\fill[black!40] (-1, 1) -- (1, 1) -- (1, -1) -- cycle;
\fill[black!40] (-1-.2, 1-.2) -- (1-.2, -1-.2) -- (-1-.2, -1-.2) -- cycle;
\draw[sstring] (1, -1) -- (-1, 1);
\draw[sstring] (1-.2, -1-.2) -- (-1-.2, 1-.2);
};
    \end{align*}

\item Perform a new SVD with truncaton dimension $k\chi$:
\[
\diagram[.8]{
\draw (-1, -1) rectangle (1, 1);
\draw (-1, 1) -- (-1.5, 1.5);
\draw (1, 1) -- (1.5, 1.5);
\draw (1, -1) -- (1.5,-1.5);
\draw (-1, -1) -- (-1.5, -1.5);
\filldraw[fill = black] (1, 1) circle (.2);
\filldraw[fill = black] (-1, -1) circle (.2);
\filldraw[fill = white] (-1, 1) circle (.2);
\filldraw[fill = white] (1, -1) circle (.2);
}\approx
\diagram[.8]{
\draw[ultra thick] (0,.5) -- node[right]{$k\chi$}(0,-.5);
\draw (0,.5) -- (1, 1.5);
\draw (0,.5) -- (-1, 1.5);
\draw (0,-.5) -- (-1, -1.5);
\draw (0,-.5) -- (1, -1.5);
\filldraw[fill = black] (0,.5-.2) arc (-90:90:.2) -- cycle;
\filldraw[fill = white] (0,.5-.2) arc (-90:-270:.2) -- cycle;
\filldraw[fill = white] (0,-.5-.2) arc (-90:90:.2) -- cycle;
\filldraw[fill = black] (0,-.5-.2) arc (-90:-270:.2) -- cycle;
}
\Leftrightarrow
\diagram[1.3]{
\fill[black!20] (-1,0) -- (0,0) -- (0,1) -- cycle;
\fill[black!20] (1,0) -- (0,0) -- (0,-1) -- cycle;
\fill[black!40] (-1,0) -- (0,0) -- (0,-1) -- cycle;
\fill[black!40] (1,0) -- (0,0) -- (0,1) -- cycle;
} \approx
\diagram[1.3]{
\fill[black!20] (-1,0) -- (0,0) -- (0,1) -- cycle;
\fill[black!20] (1,-.2) -- (0,-.2) -- (0,-1-.2) -- cycle;
\fill[black!40] (-1,-.2) -- (0,-.2) -- (0,-1-.2) -- cycle;
\fill[black!40] (1,0) -- (0,0) -- (0,1) -- cycle;
\draw[sstring] (-1, 0) -- (1, 0);
\draw[sstring] (-1,-.2) -- (1, -.2);
};
\]
\item Contract to form a new rank-4 tensor
\[
\diagram[1.2]{
\draw[ultra thick] (0,1) -- (0,-1);
\draw (-1,0) -- (1,0);
\fill[black!40] (0,0) circle (0.1);
} = 
\diagram[1.2]{
\draw (-1, 0) -- (-.5, .5) -- (.5,-.5) -- (1, 0);
\draw[ultra thick] (-.5,.5) -- (-.5,1);
\draw[ultra thick] (.5,-.5) -- (.5, -1);
\filldraw[fill = black] (.5,-.5-.1) arc (-90:90:.1) -- cycle;
\filldraw[fill = white] (.5,-.5-.1) arc (-90:-270:.1) -- cycle;
\filldraw[fill = white] (-.5,.5-.1) arc (-90:90:.1) -- cycle;
\filldraw[fill = black] (-.5,.5-.1) arc (-90:-270:.1) -- cycle;
}\Leftrightarrow
\diagram[1.2]{
\fill[black!40] (-1, 1) -- (0, 1) -- (0,0) -- cycle;
\fill[black!40] (1, 0) -- (1, 1) -- (2,0) -- cycle;
\fill[black!20] (0,0) rectangle (1, 1);
\draw[dashed] (0,0) -- (1, 1);
}.
\]
\end{enumerate}
Using a larger environment during the two approximations can further improve the accuracy.

Then using this new rank-4 tensor, we form the transfer matrix Eq.~\eqref{eq:two gates}
\[
\diagram[1.2]{
\draw[rounded corners = 5pt] (1, 0) -- (1, .5) -- (3, 2.5);
\draw[ultra thick,rounded corners = 5pt] (2, 0) -- (2, .5) -- (0, 2.5);
\draw[knot, rounded corners = 5pt] (3, 0) -- (3, .5) -- (4, 1.5) -- (0,1.5) -- (1, 2.5);
\draw[ultra thick,knot, rounded corners = 5pt] (0,0) -- (0,.5) -- (4,.5) -- (2, 2.5);
\fill[black!40] (1.5, 1) circle (.1);
\fill[black!40] (3.5, 1) circle (.1);
\fill[black!40] (.5, 2) circle (.1);
\fill[black!40] (2.5, 2) circle (.1);
}\Leftrightarrow
\diagram[1]{
\fill[black!20] (0,0) rectangle (1, 1);
\fill[black!20] (-1, 1) rectangle (0,2);
\fill[black!20] (2, 0) rectangle (3, 1);
\fill[black!20] (3, -1) rectangle (4, 0);
\fill[black!40] (1, 0) rectangle (2, 1);
\fill[black!40] (-2, 2) -- (0,0) -- (0,1) -- (-1, 1) -- (-1, 2) -- cycle;
\fill[black!40] (0,2) -- (0,1) -- (1, 1) -- cycle;
\fill[black!40] (2,0) -- (3,0) -- (3, -1) -- cycle;
\fill[black!40] (3,0) -- (4,0) -- (4, -1) -- (5, -1) -- (3, 1) -- cycle;
\draw[dashed] (2, 0) -- (1, 1);
\draw[dashed] (-3, 1) -- (-1, 1);
\draw[dashed] (-3, 1) -- (-2, 2) -- (4, -0) -- (3, -1) -- cycle;
\draw[ssstring] (4, -.5) -- (-2, 1.5);
\draw[ssstring] (0-.2,2-.2) -- (-1-.2 ,1-.2);
\draw[ssstring] (3-.2,1-.2) -- (2-.2, -.2);
}.
\]
Using the jigsaw trick, we find the geometry of this transfer matrix is
\[
\left[\frac{4}{\sqrt{10}}, 2\sqrt{10}, \frac{2}{\sqrt{10}}\right]
\]
with a moderate computational complexity $\cO(4k^3\chi^6)$. In practice, we choose $k = 2$.\\\\
Calculation of the CFT data explained above can be accessed by:\\
\includegraphics[page=10]{figures/TNRKit-paper-codesnippets.pdf}\\
Changing the geometrical shape is straightforward as\\
\includegraphics[page=8]{figures/TNRKit-paper-codesnippets.pdf}\\

Where the larger two shapes require some intermediate truncation. \texttt{trunc1} serves as a truncation of the SVD used to make 3-leg tensors out of the fixed-point 4-leg tensors (\(\chi\) in step 1 and 2 above). \texttt{trunc2} serves as a truncation used in an entanglement filtering step applied to the transfer matrix.
Other geometries can also be generated in the same fashion and can also be generalized to non-square lattices.

\section{Benchmarks}
\label{sec:benchmarks}
To validate the implementation and assess the numerical performance of the package, we present three types of benchmarks. The accuracy of the free energy for the classical Ising model, the conformal field theory (CFT) spectrum extracted throughout coarse graining, and the reproduction of selected results from the literature. Together, these tests serve as a reference for the user to make informed decisions about the methods they use.

\subsection{Ising Free Energy}
Because the 2D classical Ising model is exactly solvable, we can compare the free energy density generated by our different schemes to the exact solution \cite{Onsager1944}.

\begin{figure}
    \centering
    \includegraphics[width = \textwidth]{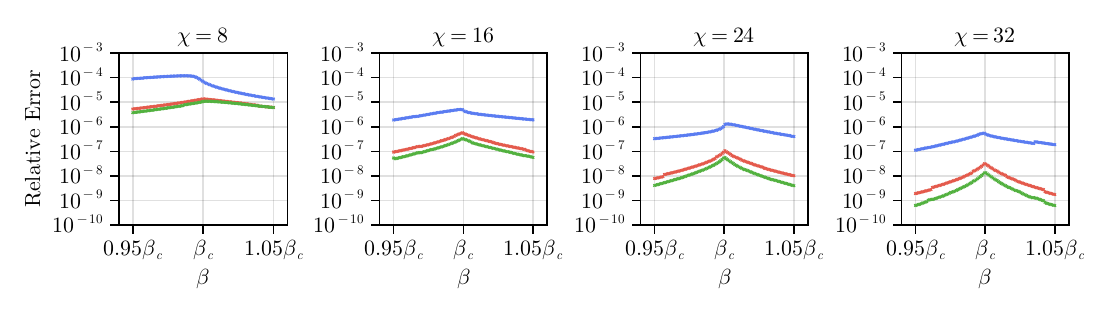}
    \caption{Accuracy vs inverse temperature for TRG (blue), HOTRG (red) and BTRG (green)  at \(\chi = 8,~16,~24 \text{ and } 32\).}
    \label{fig:accuracy-vs-beta}
\end{figure}

In Figure \ref{fig:accuracy-vs-beta} we show the relative error to Onsager's solution in function of the inverse temperature \(\beta\) at different bond-dimensions for some of the methods implemented in TNRKit.jl.\\
The main takeaway here should be that Bond-Weighted TRG is very accurate for calculating free energies. This combined with the fact that it's computational cost is the same as that of TRG -- very cheap -- makes it an excellent choice for contracting 2d tensor networks.

\begin{figure}[tb]
    \centering
    \includegraphics[width=\linewidth]{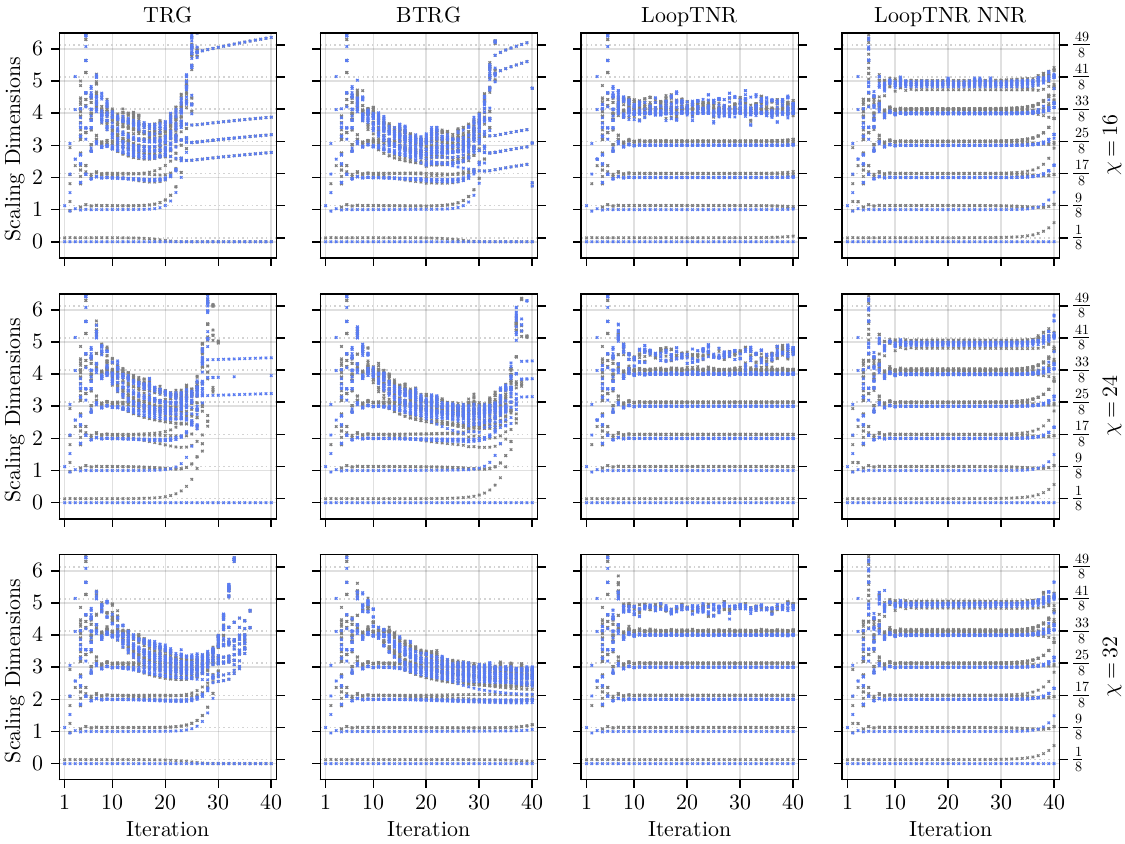}
    \caption{CFT spectrum throughout coarse graining for TRG, BTRG, LoopTNR and LoopTNR with Nuclear Norm Regularisation for $\chi = 16, 24 \text{ and } 32$. Blue data points correspond to the \(\mathbb{Z}_2\) even sector, gray data points to the \(\mathbb{Z}_2\) odd sector.}
    \label{fig:cft-data-all}
\end{figure}
\subsection{Conformal Field Theory data}
A central tool of TNR methods is the ability to calculate CFT data. TNRKit provides the user with methods to calculate the central charge, scaling dimensions and even conformal spins.

In Figure \ref{fig:cft-data-all} we show the scaling dimensions, extracted using the methods explained in section \ref{sec:cft} with modular shape \([\sqrt{2}, 2\sqrt{2}, 0]\), throughout coarse graining for different methods and bond-dimensions.
We can clearly see that, at the same computational cost, BTRG also outperforms TRG when comparing their CFT data spectra. Its lower-lying scaling dimensions stay stable for a larger amount of coarse graining steps. Both LoopTNR methods have incredibly stable spectra, and accurate data for high-lying descendants. The nuclear norm regularized LoopTNR method can provide stable data for higher laying descendants than the regular LoopTNR implementation. It must be said that getting the spectrum to show high laying scaling dimension at a high resolution takes some tuning of the hyperparameters associated with the ADMM optimisation central to LoopTNR with nuclear norm regularization.
We must note that calculating the CFT data with the \([\sqrt{2}, 2\sqrt{2}, 0]\) shape helps resolve higher-level scaling dimensions. Using a bigger patch, like one with shape  \([1, 4, 1]\), for example, would allow us to resolve even higher-level scaling dimensions. Calculations of CFT data with large patches of the network are computationally expensive however, so keep that in mind when choosing which method to use to calculate CFT data.

\subsection{Six-vertex model}
\begin{figure}[tb]
    \centering
    \includegraphics[width=0.5\linewidth]{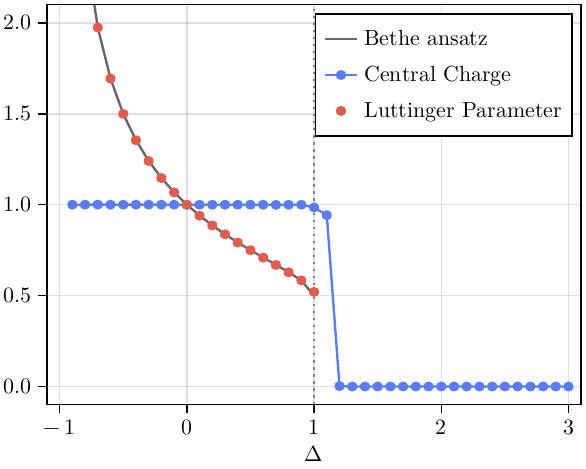}
    \caption{The central charge and Luttinger parameter of the six-vertex model obtained from LoopTNR with $\chi=40$. The black line is the exact Luttinger parameter from the Bethe ansatz. The parameters are fixed as $a=b=1$ such that the single parameter $\Delta = \frac{c^2}{2}-1$ coincides with that of the XXZ chain.}
    \label{fig:6vertex}
\end{figure}
\begin{figure}
    \centering
    \includegraphics[width = \textwidth]{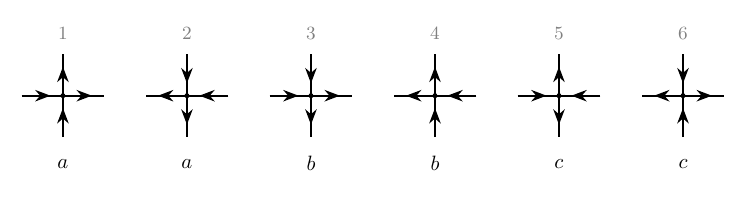}
    \caption{The ice rules for the six-vertex model.}
    \label{fig:sixvertexicerules}
\end{figure}

Next, we benchmark our method on another exactly solvable model: the six-vertex model. The partition function is defined as a product of local weights associated with the vertices, depending on three parameters $(a,b,c)$. The weights are determined by the configurations of four arrows surrounding each vertex, with arrows pointing either in or out. If one restricts to the allowed two-in two-out configurations -- the so-called ice rules -- there are precisely six such vertices, which are show in Figure~\ref{fig:sixvertexicerules}.
\begin{equation}
Z(a,b,c)=\sum_{\{\text{arrow configurations}\}} \prod_{v} w_v(a,b,c),
\end{equation}
where $w_v$ is one of the six allowed six-vertex weights. Using the standard notation,
\begin{equation}
w_1=w_2=a,\qquad
w_3=w_4=b,\qquad
w_5=w_6=c.
\end{equation}
It is known that there is a one-to-one correspondence with this model and the quantum XXZ chains with $\Delta = \frac{2c^2-a^2-b^2}{2ab}$.

The phase diagram is completely characterized by $\Delta$. In particular, the gapless Tomonaga--Luttinger-liquid(TLL)~\cite{Tomonaga:1950zz, Luttinger:1963zz} regime lies in
\begin{equation}
-1\leq\Delta\leq1,
\end{equation}
while ordered phases appear outside this window. In the tensor network representation, the partition function is the contraction of a single tensor, in accordance with the ice rules of Fig.~\ref{fig:sixvertexicerules}:
\begin{align}\label{eq:tensor_sixvertex}
    T_{0000} = T_{1111} = a,\nonumber\\
    T_{0110} = T_{1001} = b,\nonumber\\
    T_{1010} = T_{0101} = c.
\end{align}
We benchmark this model from two rather different angles: the central charge and the Luttinger parameter. The critical phase, $-1 \leq \Delta \leq 1$, is described by a $c=1$ conformal field theory, not as a single isolated point, but as a one-parameter family labeled by the Luttinger parameter. The central charge is useful because it tells us where the critical phase ends: numerically, it lets us distinguish the Tomonaga-Luttinger liquid phase from the gapped phase, since the latter should give $c=0$. The Luttinger parameter then goes one step further, telling us which member of the critical family we are looking at. In particular, the point $\Delta=1$ is the universal Berezinskii-Kosterlitz-Thouless transition, where $K=\frac{1}{2}$. 

In the TLL phase, the vertex operators have scaling dimensions
$\Delta_{m,n} = n^2 K + \frac{m^2}{4K}.$
This means that the lowest scaling dimension,
$\Delta_1 = \frac{1}{4K},$
may be used to extract the Luttinger parameter $K$. Figure~\ref{fig:6vertex} shows the central charge and Luttinger parameter obtained from LoopTNR with $\chi=40$ at the 24th RG step. The central charge drops precisely at the phase boundary, while the Luttinger parameter is in good agreement with the exact Bethe-ansatz result~\cite{Takahashi:1999bgb,Baxter:1982zz}. 

It is worth mentioning that the numerical data for the central charge at $\Delta=1.1$ in Fig.~\ref{fig:6vertex} apparently have a non-zero value. This is due to the logarithmic finite-size effect in the vicinity of the Berezinskii-Kosterlitz-Thouless transition point at $\Delta=1.$ The data points from a finite RG step capture this effect, but are expected to flow to $c=0$ if we keep coarse-graining. In the figure above, however, we stopped at 24 RG steps to avoid numerical errors arising from finite-bond-dimension effects.

\subsection{Gross-Neveu model}
We provide a benchmark of the LoopTNR algorithm for the single-flavour Gross-Neveu model on a square lattice with a Wilson fermion discretisation at finite chemical potential $\mu$ as shown in Eq.\ref{eq:grossneveu}, taken from Ref.~\citenum{akiyama2023implementationbondweightingmethod}. Here $r$ is the Wilson Fermion discretisation parameter which we set to 1. We use an initial tensor which has fermionic \(\mathbb{Z}_2\) symmetric spaces associated with its legs. The Wilson fermion formulation explicitly breaks chiral symmetry, which simplifies the construction of the tensor network considerably. Chiral-symmetry-preserving discretisations such as staggered fermions are also accessible to TNR methods but lead to an anisotropic lattice tensor network, making the construction somewhat more involved.

\begin{align}\label{eq:grossneveu}
    S[\bar\psi, \psi] = &-\frac{1}{2} \sum_{n \in \Lambda_2} \sum_{\nu = 1,2} \left [e^{\mu \delta_{\nu,2}} \bar\psi_n(r\mathbb{1} - \gamma_\nu) \psi_{n + \nu} + e^{-\mu \delta_{\nu,2}} \bar\psi_{n + \nu}(r\mathbb{1} + \gamma_\nu) \psi_{n}  \right]\nonumber\\ 
    &+ (m + 2r) \sum_{n \in \Lambda_2} \bar\psi_n \psi_n - \frac{g^2}{2} \sum_{n \in \Lambda_2} \left[(\bar\psi_n \psi_n)^2 + (\bar\psi_n i \gamma_5 \psi_n)^2 \right]
\end{align}

\begin{figure}[ht]
    \centering
    \begin{subfigure}{0.48\textwidth}
    \centering
        \includegraphics[width=\linewidth]{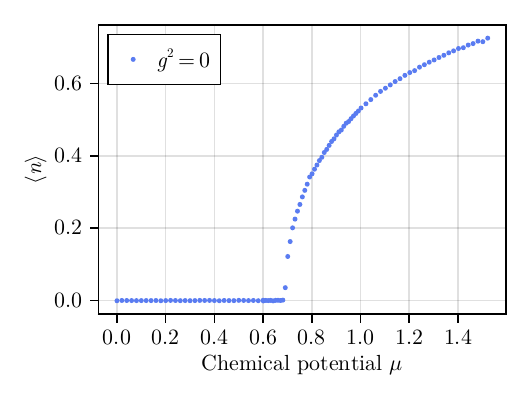}
        \caption{Number density as a function of $\mu$ at $m = -1$.}
        \label{subfig:g20}
    \end{subfigure}
    \hfill
    \begin{subfigure}{0.48\textwidth}
        \centering
        \includegraphics[width=\linewidth]{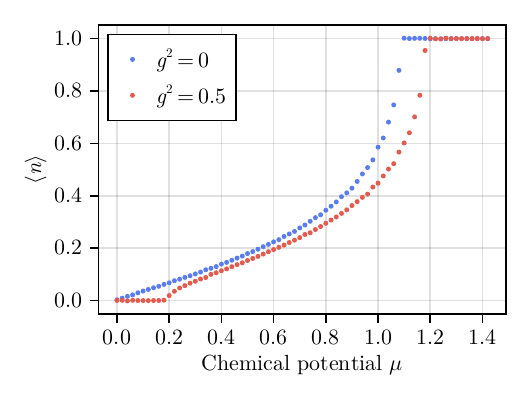}
        \caption{Number density as a function of $\mu$ at $m = 0$.}
    \end{subfigure}
        \caption{LoopTNR results for the single-flavour Gross-Neveu model.}
    \label{fig:Grossneveuakiyama}
\end{figure}

We compute the number density by
\begin{equation}
    \langle n \rangle = \frac{1}{V}
    \frac{\ln Z(\mu + \Delta\mu) - \ln Z(\mu - \Delta\mu)}{2\,\Delta\mu}
\end{equation}
via finite difference of the free energy with the chemical potential $\mu$ and find that the LoopTNR results at bond dimension $D = 16$ are in good agreement with the exact results for coupling $g^2 = 0$ and $g^2 = 0.5$, at bare mass $m = -1$ and at the critical point $m = 0$, as shown in Fig.~\ref{fig:Grossneveuakiyama}.

\section{Conclusion}
We have presented TNRKit, a Julia package for tensor-network renormalization-group methods, ranging from TRG, HOTRG, ATRG, and BTRG to the state-of-the-art LoopTNR and nuclear-norm-regularized LoopTNR algorithms. The package offers a unified interface to these methods, together with support for Abelian, non-Abelian, and fermionic symmetries through TensorKit~\cite{tensorkit}. In this way, it achieves substantial computational savings~\footnote{For example, 40 steps of LoopTNR for the critical Ising model at $\chi = 16$ take only 17 seconds when $\mathbb{Z}_2$ symmetry is exploited, whereas the same calculation takes 59 seconds without it on a MacBook Air.} and allows for a more natural description of the physics at hand. We have illustrated the performance of the available algorithms, including the ability to handle fermionic models. In addition, we provided a detailed manual on how to extract universal CFT data. 

\subsection*{Outlook}
To remain at the forefront of open-source TNR software, we plan to keep extending the package and making the state of the art easier to use in practice. The following are the next items on our list.

\paragraph{Impurity methods for observables}
Physical observables such as the magnetization or susceptibility are currently obtained by taking numerical derivatives of the free energy with respect to external parameters. This approach is computationally costly and numerically unreliable, as differentiation tends to amplify errors in $\ln Z$. The \emph{impurity tensor} method \cite{impuritybtrg, impurityhotrg, defect_trg} offers a more direct alternative: a single tensor encoding the desired operator is inserted at a chosen site, while the rest of the network is coarse-grained as usual. This gives direct access to one- and two-point functions, the correlation length, and the operator spectrum, without any numerical differentiation. Extending the impurity methods already implemented, and supporting a more general formalism based on automatic differentiation \cite{sugimotoAD}, would be a natural next step.

\paragraph{Extension to non-square lattices}
Currently, TNRKit only supports coarse-graining algorithms on a square lattice, which admits a natural two-site coarse-graining unit cell and a straightforward tensor network contraction scheme. Many models of physical interest lie on non-square lattices such as Honeycomb, Kagome, and Triangular lattices. Extending tensor network renormalization to these geometries is non-trivial, as the coarse-graining procedure must respect the point-group symmetry of the lattice and the unit cell structure changes non-trivially under blocking. TNRKit, however, does offer corner transfer matrix renormalization group (CTMRG) algorithms for the Triangular and the Honeycomb lattices \cite{triangularlatticectmrg, honeycombctmrg, ctmrgzoo}. 

\paragraph{Three-dimensional tensor network renormalization.}
Extending TNR methods to three dimensions would unlock a vast class of systems -- 3D statistical mechanics models, lattice gauge theories, and Euclidean quantum field theories in $2+1$ dimensions. TNRKit currently has access to simple 3D algorithms for HOTRG and ATRG, while more robust algorithms such as Thermal TNR \cite{thermal&globaltnr,demeyer2026,3dzngaugetheory} will be implemented in the future. The central obstacle, however, to 3D Tensor network algorithms is the proliferation of spurious local structures analogous to the CDLs familiar from two dimensions. In 3D, these generalise to \emph{corner triple lines} (CTLs) and \emph{edge double lines} (EDLs) \cite{lyu2026latticereflectionsymmetrytensornetworkrenormalization, 3dEF}: which would be the fixed points of any naive 3D coarse-graining scheme. Crucially, unlike 2D these local correlations grow linearly with the system size. Left unaddressed, they rapidly saturate the bond dimension with redundant entanglement, rendering the renormalization group flow uncontrolled and preventing convergence to the correct fixed-point tensor. An analogous \emph{3D entanglement filtering} algorithm capable of removing these CTLs and EDLs is essential. Unlike entanglement filtering in two dimensions, as we discussed above, however, the mechanism behind such a 3D algorithm would likely be very complicated. Nevertheless, solving this problem would represent a qualitative leap forward for the field, giving tensor network methods genuine access to 3D quantum field theories and some
of the hardest open problems in lattice field theory and condensed matter physics.

\paragraph{GPU acceleration.}
TensorKit has recently begun to support GPU acceleration. We aim to extend these new capabilities to TNRKit in future releases.

\section*{Acknowledgements}
VV, AN and AU express their appreciation to MV for allowing them to stay at his pied-à-terre at sea. CQM thanks Zhengcheng Gu, Dongyu Bao, Shunyao Yu, Yingjie Wei and Gong Cheng for helpful discussions and guidance. CQM especially thanks Yingjie Wei for sharing his unpublished benchmark result on $L = 6$ transfer matrix for calculating conformal spins. The TNRKit developers would like to thank everyone who has contributed in one way or another to the project. In particular, we thank Zhengyuan Yue and Sander De Meyer for their contributions. AU is grateful to Katharine Hyatt and Yuto Sugimoto for their collaboration in implementing GPU acceleration. In addition, we thank Vic Vander Linden, Boris De Vos, Jutho Haegeman, Jarid Piceu, and Lukas Devos for their contributions. 

\paragraph{Funding information}
V.V. was supported by the Research Foundation Flanders (FWO) under doctoral fellowship No. 1196525N. A. N. was supported by FWO doctoral fellowship (grant No. 11A8E26N). AU was supported by BOF-GOA (Grant No. BOF23/GOA/021) and by FWO Junior Postdoctoral Fellowship (grant No. 3E0.2025.0049.01). A. N. acknowledges support from the European Research Council (ERC) under the European Union’s Horizon 2020 program (grant agreement No. 101125822). We also acknowledge the generous support for computational resources from EUROHPC(EHPC-DEV-2026D03-098, EHPC-DEV-2025D12-166).
CQM is supported by the funding from Hong Kong’s Research Grants Council (RFS2324-4S02, CRF C7015-24G, CRS HKU701/24).

\begin{appendix}
\section{A relation between spectra}\label{sect:AB BA}
We review a basic relation in linear algebra between characteristic polynomials of $AB$ and $BA$:
\[
p_{AB}(\lambda):=\det(\lambda I - AB) = \det(\lambda I - BA)=:p_{BA}(\lambda),
\]
where both $A$ and $B$ are $n\times n$ matrices.
\begin{proof}
    Consider two new block matrices
    \[
    C = \begin{pmatrix}
        \lambda I & A\\
        B & I
    \end{pmatrix},\quad 
    D = \begin{pmatrix}
        I & \mathbf{0}\\
        -B & \lambda I
    \end{pmatrix},
    \]
    by
    \[
    \det(CD) = \det(DC),
    \]
    we have
    \[
    \lambda^n \det(\lambda I - AB) = \lambda^n\det(\lambda I - BA).
    \]
    Thus as polynomials,
    \[
    \det(\lambda I - AB) = \det(\lambda I - BA).
    \]
\end{proof}
Therefore, $AB$ and $BA$ have the same spectrum and degeneracies. More intuitively, if $v$ is an eigenstate of $AB$, i.e. $ABv = \lambda v$, then left multiplying by $B$ gives $BABv = \lambda Bv$. Therefore $Bv$ is an eigenstate of $BA$ with the same eigenvalue $\lambda$. The reverse direction is symmetric.
We emphasize that neither $A$ nor $B$ is required to be invertible.

\numberwithin{equation}{section}
\end{appendix}

\bibliography{SciPost_Example_BiBTeX_File.bib}


\end{document}